\documentclass[prl,12pt]{revtex4}
\usepackage{amssymb}
\usepackage{amsmath}
\usepackage{graphicx}
\usepackage{color}
\DeclareGraphicsExtensions{.jpg,.jpeg,.pdf,.png,.mps,.eps,.ps}

\newcommand{\beq}{\begin{equation}}
\newcommand{\eeq}{\end{equation}}
\newcommand{\bea}{\begin{eqnarray}}
\newcommand{\ena}{\end{eqnarray}}
\newcommand{\etal}{{\it et al.}}
\newcommand{\ie}{{\it i.e.}}
\newcommand{\eg}{{\it e.g.}}
\newcommand{\etc}{{\it etc.}}
\newcommand{\lsim}{\mathrel{\mathop{\kern 0pt \rlap
{\raise.2ex\hbox{$<$}}}
\lower.9ex\hbox{\kern-.190em $\sim$}}}
\newcommand{\gsim}{\mathrel{\mathop{\kern 0pt \rlap
{\raise.2ex\hbox{$>$}}}
\lower.9ex\hbox{\kern-.190em $\sim$}}}

\newcommand{\physics}[1]{{\tt physics/#1}}

\newcommand{\href}[2]{#1}

\definecolor{cyan}{cmyk}{1.,0.,0.,0.5}
\definecolor{magenta}{cmyk}{0.,1.,0.,0.5}
\definecolor{verdatre}{cmyk}{0.5,0.,0.5,0.5}
\definecolor{yellow}{cmyk}{0.,0.,0.2,0.0}
\definecolor{rouge}{cmyk}{0.,0.4,0.6,0.0}
\definecolor{orange}{cmyk}{0.,0.5,0.5,0.}
\definecolor{violet}{rgb}{0.5,0.,0.5}

\begin{document}

\noindent
\title{Discrete Symmetry in Relativistic Quantum Mechanics}
 \vskip 1.cm
\author{Guang-jiong Ni $^{\rm a,b}$}
\email{\ pdx01018@pdx.edu}
\affiliation{$^{\rm a}$ Department of Physics, Portland State University, Portland, OR97207, U. S. A.\\
$^{\rm b}$ Department of Physics, Fudan University, Shanghai, 200433, China}

\author{Suqing Chen $^{\rm b}$}
\email{\ suqing\_chen@yahoo.com}
\affiliation{$^{\rm b}$ Department of Physics, Fudan University, Shanghai, 200433, China}

\author{Jianjun Xu $^{\rm b}$}
\email{\ xujj@fudan.edu.cn}
\affiliation{$^{\rm b}$ Department of Physics, Fudan University, Shanghai, 200433, China}

\vskip 0.5cm
\date{\today}

\vskip 0.5cm
\begin{abstract}
EPR experiment on $K^0-\bar{K}^0$ system in 1998\cite{1} strongly hints that one should use operators $\hat{E}_c=-i\hbar\frac{\partial}{\partial t}$ and $\hat{\bf p}_c=i\hbar\nabla$ for the wavefunction (WF) of antiparticle. Further analysis on Klein-Gordon (KG) equation reveals that there is a discrete symmetry hiding in relativistic quantum mechanics (RQM) that ${\cal P}{\cal T}={\cal C}$. Here ${\cal P}{\cal T}$ means the (newly defined) combined space-time inversion (with ${\bf x}\to -{\bf x}, t\to-t$), while ${\cal C}$ the transformation of WF $\psi$ between particle and its antiparticle whose definition is just residing in the above symmetry. After combining with Feshbach-Villars (FV) dissociation of KG equation ($\psi=\phi+\chi$)\cite{2}, this discrete symmetry can be rigorously reformulated by the invariance of coupling equation of $\phi$ and $\chi$ under either the combined space-time inversion ${\cal P}{\cal T}$ or the mass inversion ($m\to -m$), which makes the KG equation a self-consistent theory. Dirac equation is also discussed accordingly. Various applications of this discrete symmetry are discussed, including the prediction of antigravity between matter and antimatter as well as the reason why we believe neutrinos are likely the tachyons. 

\end{abstract}

\maketitle \vskip 1cm

\section{I. Introduction}
\label{sec:introduction}
\setcounter{equation}{0}
\renewcommand{\theequation}{1.\arabic{equation}}

In 1956-1957, the historical discovery of the parity violation \cite{3,4} reveals that both P and C symmetries are violated to maximum in weak interactions. Then in 1964-1970, both CP and T are experimentally verified to be violated in some cases (though to a tiny degree) \cite{5,6} whereas the product symmetry CPT holds intact to this day \cite{7}. The CPT invariance in quantum field theory (QFT) was first proved by L\"{u}ders and Pauli in 1954-1957 \cite{8,9} via the introduction of the "strong reflection" for proving the CPT theorem. In 1965, Lee and Wu proposed that the definition of particle $|a\rangle$ versus its antiparticle $|\bar{a}\rangle$ should be \cite{10}
\begin{equation}\label{}
|\bar{a}\rangle=CPT|a\rangle
\end{equation}
Regrettably, the counterpart of "strong reflection" at the level of RQM went nearly unnoticed in the past decades. In this paper, we are going to study the RQM thoroughly. Not only a discrete symmetry ${\cal PT}={\cal C}$ is found in RQM as the counterpart of "strong reflection" in QFT, it is also evolved into the invariance of space-time inversion (${\bf x}\to {\bf x},t\to -t$) or mass inversion ($m\to -m$), showing that a WF in RQM is always composed of two parts in confrontation inside a particle and then RQM becomes a self-consistent theory. Furthermore, this symmetry can serve as a "theoretical tool" in searching for new applications in today's physics.

The organization of this paper is as follows: In section II, the EPR paradox \cite{11} is discussed together with the $K^0-\bar{K}^0$ correlation experimental data \cite{1}, yielding a strong hint that the energy-momentum operators for antiparticle's WF should be $\hat{E}_c=-i\hbar\frac{\partial}{\partial t}$ and $\hat{\bf p}_c=i\hbar\nabla$ respectively. Section III is focused on a discrete symmetry ${\cal P}{\cal T}={\cal C}$, here ${\cal P}{\cal T}$ means the (newly defined) combined space-time inversion (with ${\bf x}\to -{\bf x}, t\to-t$), while ${\cal C}$ the transformation of WFs between particle and antiparticle, whose definition is just residing in the symmetry. Then after combining with FV dissociation of KG equation \cite{2} in which the WF $\psi$ is composed of two fields: $\psi=\phi+\chi$, the above symmetry can be realized in terms of $\phi$ and $\chi$ rigorously via the invariance of their coupling equation either under the space-time inversion or a mass inversion ($m\to -m$). In this way, the probability density is ensured to be positive definite for WFs of either particle or antiparticle. Section IV ascribes various phenomena in the theory of special relativity (SR) to the effects of enhancement of the hidden $\chi$ field in a moving particle. In section V, Dirac equation is discussed accordingly with the importance of helicity being stressed. Section VI contains a brief discussion on the QFT. Sections VII, VIII and IX are devoting to seek for possible applications of the above symmetry in today's physical problems: why a parity violation phenomenon was overlooked since 1956-1957? Why we believe neutrinos are likely the tachyons? And the prediction of antigravity between matter and antimatter. The last section X contains a summary. In the Appendix, the Klein paradox is solved for both KG equation and Dirac equation without resorting to the "hole theory".

\section{II. What the $K^0\bar{K}^0$ correlation experimental data are telling?}
\label{sec:correlation}
\setcounter{equation}{0}
\renewcommand{\theequation}{2.\arabic{equation}}

To our knowledge, beginning from Bohm and Bell \cite{12}, physicists gradually turned their research of EPR paradox \cite{11} onto the entangled state composed of electrons, especially photons with spin and achieved fruitful results. However, as pointed out by Guan (1935-2007), EPR's paper \cite{11} is focused on two spinless particles and Guan found that there is a commutation relation hiding in such a system as follows \cite{13}:\\
Consider two particles in one dimensional space with positions $x_i\,(i=1,2)$ and momentum operators $\hat{p}_i=-i\hbar\frac{\partial}{\partial x_i}
$. Then a commutation relation arises as
\begin{equation}\label{1}
[x_1-x_2,\hat{p}_1+\hat{p}_2]=0
\end{equation}
According to QM's principle, there may be a kind of common eigenstate having eigenvalues of these two commutative (\ie, compatible)observables like:
\begin{equation}\label{2}
p_1+p_2=0,\;(p_2=-p_1)\quad \text{and}\quad (x_1-x_2)=D
\end{equation}
with $D$ being their distance. The existence of such kind of eigenstate described by Eq.(\ref{2}) puzzled Guan, he asked: "How can such kind of quantum state be realized?" A discussion between Guan and one of present authors (Ni) in 1998 led to a paper \cite{14}.

Here we are going to discuss further, showing that the correlation experiment on a $K^0\bar{K}^0$ system (which just realized an entangled state composed of two spinless particles) in 1998 by CPLEAR collaboration \cite{1} actually revealed some important features of QM and then answered the puzzle raised by EPR in a surprising way. First, besides Eq.(\ref{1}), let us consider another three commutation relations simultaneously:
\begin{equation}\label{3}
[t_1+t_2,\hat{E}_1-\hat{E}_2]=0
\end{equation}\\[-17mm]
\begin{equation}\label{4}
 [x_1+x_2,\hat{p}_1-\hat{p}_2] =0
\end{equation}
\begin{equation}\label{5}
 [t_1-t_2,\hat{E}_1+\hat{E}_2]=0
\end{equation}
($E_i=i\hbar\frac{\partial}{\partial t_i}$ with $t_i$ being the time during which the {\it i}-th particle is detected). In accordance with Ref.\cite{1}, we also focus on back-to-back events. The evolution of $K^0\bar{K}^0$'s wavefunction (WF) will be considered in three inertial frames: The center-of-mass system $S$ is at rest in laboratory with its origin $x=0$ located at the apparatus' center, where the antiprotons' beam is stopped inside a hydrogen gas target to create $K^0\bar{K}^0$ pairs by $p\bar{p}$ annihilation. The $K^0\bar{K}^0$ pairs are detected by a cylindrical tracking detector located inside a solenoid providing a magnetic field parallel to the antiprotons' beam. For back-to-back events, the space-time coordinates in Eqs.(\ref{1})-(\ref{5}) refer to particles moving to the right ($x_1>0$) and left ($x_2<0$) respectively. Second, we take an inertial system $S'$ with its origin located at particle 1 (\ie, $x'_1=0$). $S'$ is moving in a uniform velocity $v$ with respect to $S$. (For Kaon's momentum of $800\,MeV/c,\;\beta=v/c=0.849$). Another $S''$ system is chosen with its origin located at particle 2 ($x''_2=0$). $S''$ is moving in a velocity ($-v$) with respect to $S$. Thus we have Lorentz transformation among the space-time coordinates being
\begin{equation}\label{6}
\left\{
  \begin{array}{ll}
    x'=\dfrac{x-vt}{\sqrt{1-\beta^2}}, & \\[5mm]
    t'=\dfrac{t-vx/c^2}{\sqrt{1-\beta^2}}, &
  \end{array}
\right.\qquad
\left\{
  \begin{array}{ll}
    x''=\dfrac{x+vt}{\sqrt{1-\beta^2}}, & \\[5mm]
    t''=\dfrac{t+vx/c^2}{\sqrt{1-\beta^2}}, &
  \end{array}\right.
\end{equation}
Here $t'_1$ and $t''_2$ correspond to the proper time $t_a$ and $t_b$ in Ref.\cite{1} respectively. The common time origin $t=t'=t''=0$ is adopted.\\
A $K^0\bar{K}^0$ pair, created in a $J^{PC}=1^{--}$ antisymmetric state, can be described by a two-body WF depending on time as (\cite{1}, see also \cite{15,16})
\begin{equation}\label{7}\begin{array}{l}
|\Psi(0,0)\rangle^{(antisym)}=\dfrac{1}{\sqrt2}\left[|K^0(0)\rangle_a|\bar{K}^0(0)\rangle_b
-|\bar{K}^0(0)\rangle_a|K^0(0)\rangle_b\right]\\[5mm]
|\Psi(t_a,t_b)\rangle^{(antisym)}=\dfrac{1}{\sqrt2}\left[|K_S(0)\rangle_a|K_L(0)\rangle_be^{-i(\alpha_St_a+\alpha_Lt_b)}
-|K_L(0)\rangle_a|K_S(0)\rangle_be^{-i(\alpha_Lt_a+\alpha_St_b)}\right]
\end{array}\end{equation}
with
\begin{equation}\label{7b}
|K_S\rangle=\dfrac{1}{\sqrt2}[|K^0\rangle-|\bar{K}^0\rangle],\;|K_L\rangle=\dfrac{1}{\sqrt2}[|K^0\rangle+|\bar{K}^0\rangle]
\end{equation}
where the CP violation has been neglected and $\alpha_{S,L}=m_{S,L}-i\gamma_{S,L}/2$, $m_{S,L}$ and $\gamma_{S,L}$ being the $K_{S,L}$ masses and decay widths, respectively. From Eq.(\ref{7}), the intensities of events with like-strangeness ($K^0K^0$ or $\bar{K}^0\bar{K}^0$) and unlike-strangeness ($K^0\bar{K}^0$ or $\bar{K}^0K^0$) can be evaluated as
\begin{equation}\label{8}
I_{like}^{(antisy)}(t_a,t_b)=\dfrac{1}{8}e^{-2\gamma\tilde{t}}\left\{e^{-\gamma_S|t_a-t_b|}+e^{-\gamma_L|t_a-t_b|}
-2e^{-\gamma|t_a-t_b|}\cos[\Delta m(t_a-t_b)]\right\}
\end{equation}
\begin{equation}\label{9}
I_{unlike}^{(antisy)}(t_a,t_b)=\dfrac{1}{8}e^{-2\gamma\tilde{t}}\left\{e^{-\gamma_S|t_a-t_b|}+e^{-\gamma_L|t_a-t_b|}
+2e^{-\gamma|t_a-t_b|}\cos[\Delta m(t_a-t_b)]\right\}
\end{equation}
where $\Delta m=m_L-m_S,\;\gamma=(\gamma_S+\gamma_L)/2$ and $\tilde{t}=t_a\,(\text{for}\;t_a<t_b)$ or $\tilde{t}=t_b\,(\text{for}\;t_a>t_b)$.\\
Similarly, for $K^0\bar{K}^0$ created in a $J^{PC}=0^{++}$ or $2^{++}$ symmetric state as:
\begin{equation}\label{10}\begin{array}{l}
|\Psi(0,0)\rangle^{(sym)}=\dfrac{1}{\sqrt2}\left[|K^0(0)\rangle_a|\bar{K}^0(0)\rangle_b
+|\bar{K}^0(0)\rangle_a|K^0(0)\rangle_b\right]\\[5mm]
|\Psi(t_a,t_b)\rangle^{(sym)}=\dfrac{1}{\sqrt2}\left[|K_L(0)\rangle_a|K_L(0)\rangle_be^{-i(\alpha_Lt_a+\alpha_Lt_b)}
-|K_S(0)\rangle_a|K_S(0)\rangle_be^{-i(\alpha_St_a+\alpha_St_b)}\right]
\end{array}\end{equation}
the predicted intensities read
\begin{equation}\label{11}\begin{array}{l}
I_{like}^{(sym)}(t_a,t_b)=\dfrac{1}{8}\left\{e^{-\gamma_S(t_a+t_b)}+e^{-\gamma_L(t_a+t_b)}
-2e^{-\gamma(t_a+t_b)}\cos[\Delta m(t_a+t_b)]\right\}\\[5mm]
I_{unlike}^{(sym)}(t_a,t_b)=\dfrac{1}{8}\left\{e^{-\gamma_S(t_a+t_b)}+e^{-\gamma_L(t_a+t_b)}
+2e^{-\gamma(t_a+t_b)}\cos[\Delta m(t_a+t_b)]\right\}
\end{array}\end{equation}
The experiment \cite{1} reveals that the $K^0\bar{K}^0$ pairs are mainly created in the antisymmetric state shown by Eqs.(\ref{8})-(\ref{9}) while the contribution in a symmetric state shown by Eqs.(\ref{10})-(\ref{11}) accounts for $7.4\%$.\\
What we learn from Ref.\cite{1} in combination with Eqs.(\ref{1})-(\ref{5}) are as follows:

(a)\ Because only back-to-back events are involved in the $S$ system, we denote three commutative operators as: the "distance" operator $\hat{D}=x_1-x_2=v(t_1+t_2)$, $\hat{A}=\hat{p}_1+\hat{p}_2$ and $\hat{B}=\hat{E}_1-\hat{E}_2$, Eqs.(\ref{1}) and (\ref{3}) read
\begin{equation}\label{12}
[\hat{D},\hat{A}]=0,\;[\hat{D},\hat{B}]=0,\;[\hat{A},\hat{B}]=0
\end{equation}
So they may have a kind of common eigenstate during the measurement composed of $K^0K^0$ and projected from the symmetric state shown by Eq.(\ref{10}). It is assigned by a continuous eigenvalue $D_j=v(t_1+t_2)$ (with continuous index $j$) of operator $\hat{D}$ acting on the WF, $\Psi^{sym}_{K^0K^0}(x_1,t_1;x_2,t_2)$, as\footnotemark[1]\footnotetext[1]{The WF reads approximately as:
\begin{equation*}
\Psi^{sym}_{K^0K^0}(x_1,t_1;x_2,t_2)\sim e^{i(p_1x_1-E_1t_1)}e^{i(p_2x_2-E_2t_2)}\eqno{(2.14b)}
\end{equation*}
which can be calculated from $\langle K^0K^0|\Psi(t_a,t_b)\rangle^{sym}$ with two terms. The squares of WF's  amplitude reproduces the $I_{like}^{(sym)}(t_a,t_b)$ in Eq.(\ref{11}).}
\begin{equation*}\label{13}
\hat{D}\Psi^{sym}_{K^0K^0}(x_1,t_1;x_2,t_2)=D_j\Psi^{sym}_{K^0K^0}(x_1,t_1;x_2,t_2)
=v(t_1+t_2)\Psi^{sym}_{K^0K^0}(x_1,t_1;x_2,t_2)\eqno{(2.14a)}
\end{equation*}\\[-14mm]
\setcounter{equation}{14}
\begin{equation}\label{14}
\hat{A}\Psi^{sym}_{K^0K^0}(x_1,t_1;x_2,t_2)=A^{like}_j\Psi^{sym}_{K^0K^0}(x_1,t_1;x_2,t_2)=(p_1+p_2)\Psi^{sym}_{K^0K^0}(x_1,t_1;x_2,t_2)
\end{equation}
\begin{equation}\label{15}
\hat{B}\Psi^{sym}_{K^0K^0}(x_1,t_1;x_2,t_2)=B^{like}_j\Psi^{sym}_{K^0K^0}(x_1,t_1;x_2,t_2)=(E_1-E_2)\Psi^{sym}_{K^0K^0}(x_1,t_1;x_2,t_2)
\end{equation}
where the lowest eigenvalue of $\hat{A}$ is $A^{like}_j=p_1+p_2=0,\,(p_2=-p_1)$, and that of $\hat{B}$ is $B^{like}_j=E_1-E_2=0,\,(E_2=E_1)$ respectively. These eigenstates of like-strangeness events predicted by Eq.(\ref{10}) are really observed in the experiment \cite{1} (these eigenstates of $K^0K^0$ were overlooked in the Ref.\cite{14}).

(b)\ The more interesting case occurs for $K^0\bar{K}^0$ pair created in the antisymmetric state with intensity given by Eq.(\ref{9}) being a function of $(t_a-t_b)$ (not $(t_a+t_b)$ as shown by Eq.(12) for symmetric states) which is proportional to $(t_1-t_2)$ in the $S$ system. In the EPR limit $t_1=t_2$, $K^0\bar{K}^0$ events dominate whereas like-strangeness events are strongly suppressed as shown by Eq.(\ref{8}) (see Fig.1 in \cite{1}). So the experimental facts remind us of the possibility that $K^0\bar{K}^0$ events may be related to common lowest (zero) eigenvalues of some commutative operators (just like what happened in Eqs.(\ref{14}) and (\ref{15}) for operators $\hat{A}$ and $\hat{B}$ (which are applied to symmetric states (due to $\hat{D}=x_1-x_2=v(t_1+t_2)$) but are not suitable for antisymmetric states), there are another three operators shown by Eqs.(\ref{4}) and (\ref{5}) being:
the operator of "flight-path difference" $\hat{F}=x_1+x_2=v(t_1-t_2)$, $\hat{M}=\hat{p}_1-\hat{p}_2$ and $\hat{G}=\hat{E}_1+\hat{E}_2$ with commutation relations as:
\begin{equation}\label{16}
[\hat{F},\hat{M}]=0,\;[\hat{F},\hat{G}]=0,\;[\hat{M},\hat{G}]=0
\end{equation}
which are just suitable for antisymmetric states. For $K^0\bar{K}^0$ back-to-back events, assume that one of two particles, say 2, is an antiparticle with its momentum and energy operators being
\begin{equation}\label{17}
\hat{p}_x^c=i\hbar\dfrac{\partial}{\partial x},\;\hat{E}^c=-i\hbar\dfrac{\partial}{\partial t}
\end{equation}
(the superscript $c$ means "antiparticle") versus that for particle being
\begin{equation}\label{17-1}
\hat{p}_x=-i\hbar\dfrac{\partial}{\partial x},\;\hat{E}=i\hbar\dfrac{\partial}{\partial t}
\end{equation}
For instance, a freely moving particle's WF reads\footnotemark[1]\footnotetext[1]{Please see the derivation of Eqs.(2.20) and (2.21) from the quantum field theory (QFT) at the end of section VI.}:
\begin{equation}\label{18}
\psi(x,t)\sim\exp\left[\frac{i}{\hbar}(px-Et)\right]
\end{equation}
whereas
\begin{equation}\label{19}
\psi_c(x,t)\sim\exp\left[-\frac{i}{\hbar}(p_cx-E_ct)\right]
\end{equation}
for its antiparticle with $p_c=p$ and $E_c\,(>0)$ being momentum and energy of the antiparticle in accordance with Eq.(2.18). If using Eqs.(2.18)-(\ref{19}), we find
\begin{equation}\label{20}
\hat{F}\Psi^{antisym}_{K^0\bar{K}^0}(x_1,t_1;x_2,t_2)=F^{unlike}_k\Psi^{antisym}_{K^0\bar{K}^0}(x_1,t_1;x_2,t_2)=v(t_1-t_2)\Psi^{antisym}_{K^0\bar{K}^0}(x_1,t_1;x_2,t_2)
\end{equation}
with continuous index $k$ referring to continuous eigenvalues $F_k=v(t_1-t_2)$. Here, the WF in space-time of this system during measurement reads approximately:
\begin{equation}\label{21}
\Psi^{antisym}_{K^0\bar{K}^0}(x_1,t_1;x_2,t_2)\sim e^{i(p_1x_1-E_1t_1)}e^{-i(p_2^cx_2-E_2^ct_2)}
\end{equation}
with antiparticle 2 moving opposite to particle 1 and $p_2^c=-p_1$.

Now we use $\hat{M}(=\hat{p}_1-\hat{p}_2)=\hat{p}_1+\hat{p}^c_2$ on $K^0\bar{K}^0$ system, yielding
\begin{equation}\label{22}
\hat{M}\Psi^{antisym}_{K^0\bar{K}^0}(x_1,t_1;x_2,t_2)=M^{unlike}_k\Psi^{antisym}_{K^0\bar{K}^0}(x_1,t_1;x_2,t_2)
=(p_1+p^c_2)\Psi^{antisym}_{K^0\bar{K}^0}(x_1,t_1;x_2,t_2)
\end{equation}

Similarly, we have $\hat{G}(=\hat{E}_1+\hat{E}_2)=\hat{E}_1-\hat{E}^c_2$ and find
\begin{equation}\label{223}
\hat{G}\Psi^{antisym}_{K^0\bar{K}^0}(x_1,t_1;x_2,t_2)=G^{unlike}_k\Psi^{antisym}_{K^0\bar{K}^0}(x_1,t_1;x_2,t_2)
=(E_1-E^c_2)\Psi^{antisym}_{K^0\bar{K}^0}(x_1,t_1;x_2,t_2)
\end{equation}
Hence we see that once Eqs.(2.18) and (\ref{19}) are accepted, the WFs $\Psi^{antisym}_{K^0\bar{K}^0}(x_1,t_1;x_2,t_2)$ show up in experiments as the only WFs with strongest intensity at the EPR limit ($t_1=t_2$) corresponding to their three eigenvalues being all zero: $F_k=M^{unlike}_k=G^{unlike}_k=0$ and they won't change even when accelerator's energies are going up.\\
If using Eq.(2.18), the eigenvalues of $\hat{A}$ and $\hat{B}$ for the WF $\Psi^{antisym}_{K^0\bar{K}^0}(x_1,t_1;x_2,t_2)$ are $A^{unlike}_j=p_1-p^c_2=2p_1$ and $B^{unlike}_j=E_1+E^c_2=2E_1$ respectively, while that of $\hat{M}$ and $\hat{G}$ for the WF $\Psi^{antisym}_{K^0K^0}(x_1,t_1;x_2,t_2)$ are $M^{like}_k=p_1-p_2=2p_1$ and $G^{like}_k=E_1+E_2=2E_1$, respectively, those eigenvalues are much higher than zero and going up with the accelerator's energy.

Something is very interesting here: If we deny Eq.(2.18) but insist on unified operators $\hat{p}$ and $\hat{E}$ for both particle and antiparticle, there would be no difference in eigenvalues between like-strangeness events and unlike-strangeness ones. For example, the $M^{unlike}_k$ and $G^{unlike}_k$ would be $2p_1$ and $2E_1$ too (instead of "0" as in Eqs.(\ref{22}) and (2.25)). This would mean that three commutative operators $\hat{F},\hat{M}$ and $\hat{G}$ are not enough to distinguish the WF $\Psi^{antisym}_{K^0\bar{K}^0}(x_1,t_1;x_2,t_2)$ from the WF $\Psi^{antisym}_{K^0K^0}(x_1,t_1;x_2,t_2)$ even they behave so differently as shown by Eqs.(\ref{8}) and (\ref{9})), especially at the EPR limit ($t_1=t_2$).\\
Eq.(2.18) together with the identification of WF $\Psi^{antisym}_{K^0\bar{K}^0}(x_1,t_1;x_2,t_2)$ by three zero eigenvalues implies that the difference of a particle from its antiparticle is not something hiding in the "intrinsic space" like opposite charge (for electron and positron) or opposite strangeness (for $K^0$ and $\bar{K}^0$) but can be displayed in their WFs evolving in space-time at the level of QM.

In summary, instead of one set of WF with its operators (Eqs.(2.20) and (2.19)), two sets of WFs with operators separately (shown as Eqs.(2.18)-(2.21)) are strongly supported by the original EPR paradox and its "solution" provided by the $K^0-\bar{K}^0$ experiment.

To our knowledge, Eq.(2.18) can be found at a page note of a paper by Konopinski and Mahmaud in 1953 \cite{17}, also appears in Refs.\cite{14,18,19,20,21,22,23}.

\section{III. How to Make Klein-Gordon Equation a Self-Consistent Theory in RQM ? A Discrete Symmetry.}
\label{sec:invariance}
\setcounter{equation}{0}
\renewcommand{\theequation}{3.\arabic{equation}}

\subsection{IIIA. The negative energy solution and the WF of antiparticle}
\setcounter{equation}{0}
\renewcommand{\theequation}{3.\arabic{equation}}

Let us begin with the energy conservation law for a particle in classical mechanics:
\begin{equation}\label{44}
E=\frac{1}{2m}{\bf p}^2+V({\bf x})
\end{equation}
Consider the rule promoting observables into operators:
\begin{equation}\label{45}
E\to\hat{E}=i\hbar\frac{\partial}{\partial t},\quad {\bf p}\to \hat{\bf p}=-i\hbar\nabla
\end{equation}
and let Eq.(\ref{44}) act on a wavefunction (WF) $\psi({\bf x},t)$, the Schr\"{o}dinger equation
\begin{equation}\label{46}
i\hbar\frac{\partial}{\partial t}\psi({\bf x},t)=-\frac{\hbar^2}{2m}\nabla^2\psi({\bf x},t)+V({\bf x})\psi({\bf x},t)   \end{equation}
follows immediately. In mid 1920's, considering the kinematical relation for a particle in the theory of special relativity (SR):
\begin{equation}\label{47}
(E-V)^2=c^2{\bf p}^2+m^2c^4
\end{equation}
and using Eq.(\ref{45}) again, the Klein-Gordon (KG) equation was established as:
\begin{equation}\label{48}
(i\hbar\frac{\partial}{\partial t}-V)^2\psi({\bf x},t)=-c^2\hbar^2\nabla^2\psi({\bf x},t)+m^2c^4\psi({\bf x},t)    \end{equation}
For a free KG particle, its plane-wave solution reads:
\begin{equation}\label{49}
\psi({\bf x},t)\sim \exp[\frac{i}{\hbar}({\bf p}\cdot{\bf x}-Et)]
\end{equation}
However, two difficulties arose:

(a) The energy $E$ in Eq.(\ref{49}) has two eigenvalues:
\begin{equation}\label{50}
E=\pm\sqrt{c^2{\bf p}^2+m^2c^4}
\end{equation}
In general, $V\neq0$, the WFs of KG particle's energy eigenstates can always be divided into two parts:
\begin{eqnarray}
  \psi &\sim & \exp(-\frac{i}{\hbar}Et),\quad E>0 \\[2mm]
  \psi &\sim & \exp(-\frac{i}{\hbar}Et),\quad E<0
\end{eqnarray}
where only the original operators Eq.(3.2) are used. But what the "negative energy" means?

(b)The continuity equation is derived from Eq.(\ref{48}) as
\begin{equation}\label{51}
\frac{\partial\rho}{\partial t}+\nabla\cdot{\bf j}=0
\end{equation}
where
\begin{equation}\label{52}
\rho=\frac{i\hbar}{2mc^2}(\psi^*\frac{\partial}{\partial t}\psi-\psi\frac{\partial}{\partial t}\psi^*)-\frac{1}{mc^2}V\psi^*\psi
\end{equation}
and
\begin{equation}\label{53}
{\bf j}=\frac{i\hbar}{2m}(\psi\nabla\psi^*-\psi^*\nabla\psi)
\end{equation}
are the "probability density" and "probability current density" respectively. While the latter is the same as that derived from Eq.(\ref{46}), Eq.(3.11) seems not positive definite and dramatically different from $\rho=\psi^*\psi$ in Eq.(\ref{46}). Why?

In hindsight, for a linear equation in RQM, either KG or Dirac equation, the emergence of WFs with both positive and negative energy ($E$) is inevitable and natural. From mathematical point of view, the set of WFs cannot be complete if without taking the negative energy solutions into account. And physicists believe that these negative-energy solutions might be relevant to antiparticles. However, we physicists admit that both a rest particle's energy $E=mc^2$ and a rest antiparticle's energy $E_c=m_cc^2=mc^2$ are positive, as verified by numerous experiments like that of pair-creation process $\gamma\to e^++e^-$. The above contradiction constructs so-called "negative-energy paradox" in RQM. For Dirac particle, majority (not all) of physicists accept the "hole theory" to explain the "paradox". But for KG particle, no such kind of "hole theory" can be acceptable. It was this "negative-energy paradox" as well as the four "commutation relations", Eqs.(\ref{1})-(\ref{5}), hidden in the two-particle system discussed by EPR \cite{11} gradually prompted us to realize that the root cause of difficulty in RQM lies in an a priori notion --- only one kind of WF with one set of operators (like Eq.(\ref{45})) can be acceptable in QM, either for NRQM or RQM.

Once getting rid of the constraint in the above notion and introducing two sets of WFs and operators for particle and antiparticle respectively, we can identify the negative energy solution, Eq.(3.9), with the antiparticle's WF directly
\begin{equation}\label{}
\psi_c\sim \exp(\frac{i}{\hbar}E_ct),\quad E_c>0
\end{equation}
which implies an antiparticle with positive energy $E_c$ by using Eq.(2.18). This claim will be proved rigorously in the next subsection.

One may ask: When you assume the negative energy solution being the WF of antiparticle, how about the difficulty of negative probability density? Below we will see how to solve these two difficulties simultaneously and make KG equation a self-consistent theory at the level of RQM.

\subsection{IIIB. The Proof of a Discrete symmetry ${\cal P}{\cal T}={\cal C}$ for KG particle}

Let us introduce an operator of (newly defined) combined space-time inversion ${\cal P}{\cal T}$ for KG equation. It should change the space-time coordinates as
\begin{equation}\label{}
{\bf x}\to -{\bf x},\, t\to -t
\end{equation}
then accordingly
\begin{equation}\label{}
\hat{\bf p}=-i\hbar\nabla\to {\cal P}{\cal T}\hat{\bf p}({\cal P}{\cal T})^{-1}=\hat{\bf p}_c=i\hbar\nabla, \hat{E}=i\hbar\frac{\partial}{\partial t}\to{\cal P}{\cal T}\hat{E}({\cal P}{\cal T})^{-1}=\hat{E}_c=-i\hbar\frac{\partial}{\partial t}
\end{equation}
Because the antiparticle has opposite charge ($-q$) versus $q$ for particle, so
\begin{equation}\label{}
V({\bf x},t)\to{\cal P}{\cal T}V({\bf x},t)({\cal P}{\cal T})^{-1}\equiv V_c({\bf x},t)=-V({\bf x},t)
\end{equation}
When performing ${\cal P}{\cal T}$ inversion on KG equation, Eq.(3.5), from left to right, we meet eventually the WF and define the antiparticle's WF as
\begin{equation}\label{}
{\cal P}{\cal T}\psi({\bf x},t)\equiv{\cal C}\psi({\bf x},t)=\psi_c({\bf x},t)
\end{equation}
Thus KG particle's equation, Eq.(3.5), is transformed into ($\hbar=1$)
\begin{equation}\label{}
(\hat{E}_c-V_c)^2\psi_c({\bf x},t)=-c^2\nabla^2\psi_c({\bf x},t)+m^2\psi_c({\bf x},t)
\end{equation}
or
\begin{equation}\label{}
(i\frac{\partial}{\partial t}-V)^2\psi_c({\bf x},t)=-c^2\nabla^2\psi_c({\bf x},t)+m^2\psi_c({\bf x},t)
\end{equation}
which is formally the same as Eq.(3.5) though we should use $\hat{{\bf p}}_c,\hat{E}_c$ for $\psi_c({\bf x},t)$. Hence the KG equation remains invariant under the ${\cal P}{\cal T}$ operation, Eqs.(3.14)-(3.17). Notice further that Eq.(3.18) is just the "quantized" equation of the kinematical relation for an antiparticle in SR
\begin{equation}\label{}
(\hat{E}_c-V_c)^2=c^2{\bf p}_c^2+m^2c^4
\end{equation}
which is the counterpart of Eq.(3.4) for a particle.
For example, a KG particle's scattering WF $\psi({\bf x},t;E_1)\sim e^{-iE_1t}\,(E_1>m)$ is attracted by an spherically symmetric potential $V(r)<0$ and so has a positive phase-shift $\delta_1>0$ (in the , say, $S(l=0)$ state). Then physically, its antiparticle's WF $\psi_c({\bf x},t;E_1^c)\sim e^{iE_1^ct}\,(E_1^c=E_1>m)$ is repelled by the potential $V_c(r)=-V(r)>0$ and has a negative phase-shift $\delta_1^c<0$.

Note that, however, corresponding to $\psi({\bf x},t;E_1)$, there is another negative energy particle's WF $\psi({\bf x},t;-E_1)\sim e^{iE_1t}$ satisfying Eq.(3.5)
\begin{equation}\label{}
(i\frac{\partial}{\partial t}-V)^2\psi({\bf x},t;-E_1)=(-E_1-V)^2\psi({\bf x},t;-E_1)=-c^2\nabla^2\psi({\bf x},t;-E_1)+m^2\psi({\bf x},t;-E_1)
\end{equation}
whose space-time behavior is precisely the same as the antiparticle's WF $\psi_c({\bf x},t;E_1^c)\sim e^{iE_1^ct}$ with $E_1^c=E_1>m$ as shown by Eq.(3.18) since $(E_1+V)^2=(E_c-V_c)^2$. Thus, for avoiding confusion, we have
\begin{equation}\label{}
{\cal P}{\cal T}\psi({\bf x},t;E_1)=\psi({\bf x},t;-E_1)={\cal C}\psi({\bf x},t;E_1)=\psi_c({\bf x},t;E_1^c)\neq\psi(-{\bf x},-t;E_1)
\end{equation}
and
\begin{equation}\label{}
{\cal P}{\cal T}\psi({\bf x},t)={\cal C}\psi({\bf x},t)=\psi(-{\bf x},-t)=\psi_c({\bf x},t)\quad (V=0)
\end{equation}
achieving the proof of the discrete symmetry ${\cal P}{\cal T}={\cal C}$ for KG particle shown by Eq.(3.17). In summary, the "negative-energy paradox" for KG equation is solved in a physical way with following advantages:

a) By using two sets of WFs and momentum-energy operators for particle and antiparticle respectively, both particle's WF $\psi({\bf x},t)$ and antiparticle's WF $\psi_c({\bf x},t)$ have positive energies $E>0$ and $E_c>0$ respectively.

b) While satisfying the same KG equation with same potential $V(r)$ formally, $\psi({\bf x},t)$ and $\psi_c({\bf x},t)$ are actually subject to opposite "force" for particle and antiparticle respectively.

c) The space-time behavior of $\psi_c({\bf x},t;E_1^c)$ can be identified with that of a negative energy particle's WF $\psi({\bf x},t;-E_1)\, (E_1=E_1^c)$, in a one-to-one correspondence. Thus from mathematical point of view, all solutions of KG equation form a complete set including both positive and negative energy values of one operator $\hat{E}=i\frac{\partial}{\partial t}$ exactly.

By contrast, usually, aiming at finding an antiparticle'S WF, one performs the CPT transformation on a particle's WF $\psi({\bf x},t)$, yielding \cite{24,25,26}
\begin{equation}\label{}
\psi({\bf x},t)\to CPT \psi({\bf x},t)=\psi(-{\bf x},-t)
\end{equation}
whose character can also be summed up as follows:

$a'$)By using one set of WF and relevant operators for both particle and antiparticle, at the LHS of Eq.(3.24), $\psi({\bf x},t)$, and $\psi(-{\bf x},-t)$ at RHS must have opposite energies inevitably.

$b'$) By design in the C transformation, $\psi({\bf x},t)$ and $\psi(-{\bf x},-t)$ in Eq.(3.24) satisfy different equations with $V$ and $V_c=-V$ respectively. But with opposite energies, they are actually subject to the same (either attractive or repulsive) "force". So one cannot distinguish particle from antiparticle through what their WFs "feel" after the CPT transformation.

$c'$) From mathematical point of view, we should keep all negative-energy solutions for one equation. However, even facing WFs in doubled numbers, we still don't know how to choose half of them for describing particle and its antiparticle separately in physics.

But we haven't solve the difficulty of negative probability density in KG equation yet, awaiting for another enlightenment which was already there since 1958.

\subsection{IIIC. Feshbach and Villars (FV) dissociation of KG WF£º$\psi=\phi+\chi$, a reformulated symmetry between $\phi$ and $\chi$ under the space-time (or mass) inversion}

In 1958, dividing the WF into $\psi=\phi+\chi$, Feshbach and Villars \cite{2} recast Eq.(\ref{48}) into two coupled Schr\"{o}dinger-like equations as:\footnotemark[1]\footnotetext[1]{Interestingly, if ignoring the coupling between $\phi$ and $\chi$ and $V=0$ in Eq.(\ref{54}), they satisfy respectively the "two equations" written down by Schr\"{o}dinger in his 6th paper in 1926, titled "Quantisation as a problem of proper values (Part IV)" (Annalen der Physik Vol.81, No.4, 1926, p104) when he invented NRQM in the form of wave mechanics.}
\begin{equation}\label{54}
\left\{
  \begin{array}{ll}
   \left(i\hbar\dfrac{\partial}{\partial t}-V\right)\phi=mc^2\phi-\dfrac{\hbar^2}{2m}\nabla^2(\phi+\chi)    \\[4mm]
    \left(i\hbar\dfrac{\partial}{\partial t}-V\right)\chi=-mc^2\chi+\dfrac{\hbar^2}{2m}\nabla^2(\phi+\chi)
  \end{array}
\right.
\end{equation}
where
\begin{equation}\label{55}
\left\{
  \begin{array}{ll}
   \phi=\dfrac{1}{2}\left[\left(1-\dfrac{1}{mc^2}V\right)\psi+\dfrac{i\hbar}{mc^2}\dot{\psi}\right]    \\[4mm]
    \chi=\dfrac{1}{2}\left[\left(1+\dfrac{1}{mc^2}V\right)\psi-\dfrac{i\hbar}{mc^2}\dot{\psi}\right]
  \end{array}
\right.
\end{equation}
($\dot{\psi}=\frac{\partial\psi}{\partial t}$). Interestingly, the "probability density", Eq.(3.11) can be recast into a difference between two positive-definite densities \cite{14,16}:
\begin{equation}\label{56}
\rho=\phi^*\phi-\chi^*\chi
\end{equation}
while the probability current density contains interference terms between $\phi$ and $\chi$:
\begin{equation}\label{}
{\bf j}=\frac{i\hbar}{2m}[(\phi\nabla\phi^*-\phi^*\nabla\phi)+(\chi\nabla\chi^*-\chi^*\nabla\chi)
+(\phi\nabla\chi^*-\chi^*\nabla\phi)+(\chi\nabla\phi^*-\phi^*\nabla\chi)]
\end{equation}
The expression of $\rho$ as shown by Eq.(3.27) strongly hints that the ${\cal P}{\cal T}={\cal C}$ symmetry proved in the last subsection may be combined with the FV dissociation of KG equation such that the positive-definite property of $\rho$ can be ensured for both particle and antiparticle.

Indeed, after inspecting Eq.(3.25) carefully, we do find a hidden symmetry in the sense that it is invariant (in its form) under the following reformulated space-time inversion $({\bf x}\to-{\bf x},t\to-t)$, \ie, ${\cal P}{\cal T}={\cal C}$ transformation:
\begin{equation}\label{3-25}
\left\{
\begin{array}{l}
{\bf x}\to-{\bf x},t\to-t,\\
 V({\bf x},t)\to -V({\bf x},t)=V_c({\bf x},t),\\
 \psi({\bf x},t)\to {\cal P}{\cal T}\psi({\bf x},t)=\psi_c({\bf x},t),\\
 \phi({\bf x},t)\to {\cal P}{\cal T}\phi({\bf x},t)=\chi_c({\bf x},t),\\
  \chi({\bf x},t)\to {\cal P}{\cal T}\chi({\bf x},t)=\phi_c({\bf x},t)
\end{array}\right.
\end{equation}
Performing transformation Eq.(3.29) on Eq.(3.26), we find $\chi_c$ satisfying the same equation of $\chi$ and $\phi_c$ satisfying that of $\phi$. They read
\begin{equation}\label{3-20}
\left\{
\begin{array}{l}
 \chi_c=\dfrac{1}{2}\left[\left(1+\dfrac{1}{mc^2}V\right)\psi_c-\dfrac{i\hbar}{mc^2}\dot{\psi}_c\right]\\[4mm]
\phi_c=\dfrac{1}{2}\left[\left(1-\dfrac{1}{mc^2}V\right)\psi_c+\dfrac{i\hbar}{mc^2}\dot{\psi}_c\right]
\end{array}\right.
\end{equation}
Remember, for $\psi_c$, we should use operator Eq.(3.15). Accordingly, the probability density for $\psi_c$ is defined as
\begin{equation}\label{3-21}
\rho\to{\cal P}{\cal T}\rho=\rho_c=\frac{i\hbar}{2mc^2}(\psi_c\dot{\psi}_c^*-\psi_c^*\dot{\psi}_c)+\frac{1}{mc^2}V\psi_c^*\psi_c
=\chi_c^*\chi_c - \phi_c^*\phi_c
\end{equation}
Similarly, we have ($\nabla\psi\to -\nabla\psi_c$)
\begin{equation}\label{61}
{\bf j}\to{\cal P}{\cal T}{\bf j}={\bf j}_c=\frac{i\hbar}{2m}(\psi^*_c\nabla{\psi}_c-\psi_c\nabla\psi^*_c)
\end{equation}
For simplicity, consider a free KG particle ($V=0$) with WF Eq.(\ref{49}). Then $|\phi|>|\chi|$
\begin{equation}\label{3-14}
\left\{
  \begin{array}{ll}
    \phi=\dfrac{1}{2}\left(1+\dfrac{E}{mc^2}\right)\psi &  \\[4mm]
    \chi=\dfrac{1}{2}\left(1-\dfrac{E}{mc^2}\right)\psi &
  \end{array},\right.\left\{
  \begin{array}{ll}
    \rho=|\phi|^2-|\chi|^2>0 &  \\[4mm]
    {\bf j}=\dfrac{1}{m}{\bf p}|\psi|^2 &
  \end{array}
\right.
\end{equation}
But for a free ($V=0$) KG antiparticle with WF Eq.(2.21), it has $|\chi_c|>|\phi_c|$
\begin{equation}\label{}
\left\{
  \begin{array}{ll}
    \phi_c=\dfrac{1}{2}\left(1-\dfrac{E_c}{mc^2}\right)\psi_c &  \\[5mm]
    \chi_c=\dfrac{1}{2}\left(1+\dfrac{E_c}{mc^2}\right)\psi_c &
  \end{array},\right.\left\{
  \begin{array}{ll}
    \rho_c=|\chi_c|^2-|\phi_c|^2>0 &  \\[4mm]
    {\bf j}_c=\dfrac{1}{m}{\bf p}_c|\psi_c|^2 &
  \end{array}
\right.
\end{equation}
Eqs.(3.33)-(3.34) satisfy all physical conditions we need. If $V\neq0$, as long as $(E-V)>0$ for particle or $(E_c-V_c)>0$ for antiparticle, the situation remains the same. However, once $(E-V)<0$ or $(E_c-V_c)<0$, some complications would occur. For further discussion, please see the Appendix.

Therefore, we see that the reformulated space-time inversion, Eq.(3.29), reflects the underlying symmetry between a particle's WF $\psi$ and its antiparticle's WF $\psi_c$. As both $E$ and $\rho$ in $\psi$ or $E_c$ and $\rho_c$ in $\psi_c$ are positive definite, all difficulties in KG equation disappear and the latter becomes a self-consistent theory.

Moreover, instead of Eq.(3.29), a "mass inversion ($m\to-m$)" can realize the same symmetry, the invariance under a ${\cal P}{\cal T}={\cal C}$ transformation, via the following operation on Eq.(3.25):
\begin{equation}\label{3-25}
\left\{
\begin{array}{l}
m\to -m_c=-m\\
 V({\bf x},t)\to V({\bf x},t)=-V_c({\bf x},t),\\
 \psi({\bf x},t)\to \psi_c({\bf x},t),\\
 \phi({\bf x},t)\to \chi_c({\bf x},t),\\
  \chi({\bf x},t)\to \phi_c({\bf x},t)
\end{array}\right.
\end{equation}
Notice that, when $m\to-m$, we have $\hat{\bf p}\to -\hat{\bf p}_c$ and $\hat{E}\to-\hat{E}_c$, \ie~ $-i\hbar\nabla\to-i\hbar\nabla$, $i\hbar\frac{\partial}{\partial t}\to i\hbar\frac{\partial}{\partial t}$, in contrast to Eq.(3.15). \footnotemark[1]\footnotetext[1]{Here $m$ always refers to the "rest mass" also the "inertial mass" for a particle or its antiparticle, see the excellent paper by Okun in Ref.\cite{27}.}

The reason why $V\to -V$ in the space-time inversion Eq.(3.29) whereas $V\to V$ in the mass inversion Eq.(3.35) can be seen from the classical equation: The Lorentz force ${\bf F}$ on a particle exerted by an external potential $\Phi$ reads: ${\bf F}=-\nabla V=-\nabla(q\Phi)=m{\bf a}$. As the acceleration ${\bf a}$ of particle will change to $-{\bf a}$ for its antiparticle, there are two alternative explanations: either due to the inversion of charge $q\to-q$ (\ie, $V\to -V$ but keeping $m$ unchanged) or due to the inversion of mass $m\to-m$ (but keeping $V$ unchanged).

\section{IV. Reinterpretation of WF and the Relativistic Effects}
\setcounter{equation}{0}
\renewcommand{\theequation}{4.\arabic{equation}}

The success of FV's dissociation of KG equation should be ascribed to their deep insight that a unified WF $\psi$ is composed of two fields $\phi$ and $\chi$ in confrontation. Note that Eq.(\ref{54}) reduces into two equations separately for a static KG particle ($V=0,\hbar=c=1$):
\begin{equation}\label{4-1}
\left\{
\begin{array}{l}
 i\dfrac{\partial}{\partial t}\phi=m\phi,\\[4mm]
  i\dfrac{\partial}{\partial t}\chi=-m\chi
\end{array}\right.
\end{equation}
with two separated solutions being:
\begin{equation}\label{4-2}
\left\{
\begin{array}{l}
 E=m>0,\\
 \phi\sim e^{-iEt},\\
  \chi=0
\end{array}\right.,\,
\left\{
\begin{array}{l}
 E=-m=-E_c<0,E_c=m>0\\
 \chi_c\sim e^{iE_ct},\\
  \phi_c=0
\end{array}\right.
\end{equation}
Once the particle (antiparticle) is moving with the velocity, $v\neq0$, $\phi$ and $\chi$ ($\chi_c$ and $\phi_c$) couple together and the WF $\psi=\phi+\chi$ ($\psi_c=\phi_c+\chi_c$) for a free particle (antiparticle) read (in one-dimensional space)
\begin{subequations}
\begin{align}
 \psi\sim\phi\sim\chi\sim \exp[i(px-Et)],&\quad (|\phi|>|\chi|)\\
  \psi_c\sim\chi_c\sim\phi_c\sim \exp[-i(p_cx-E_ct)],&\quad (|\chi_c|>|\phi_c|)
\end{align}
\end{subequations}
($p_c=p>0,E_c=E>0$) respectively. In Eq.(4.3a), $\phi$ dominates $\chi$ $(|\phi|>|\chi|)$. By contrast, in Eq.(4.3b) it is $\chi_c$ who dominates $\phi_c$ (The status remains the same for $V\neq0$ cases as discussed in the last section).

Despite $\phi$ and $\phi_c$ ($\chi$ and $\chi_c$) having the "intrinsic tendency" to evolve as $\exp[i(px-Et)]$ ($\exp[-i(px-Et)]$), however, in a WF of particle (antiparticle), $\chi(\phi_c)$ must follow $\phi(\chi_c)$ to evolve like that shown by Eq.(4.3a) (Eq.(4.3b)), as $|\phi|>|\chi|(|\chi_c|>|\phi_c|)$. So it seems suitable to name $\phi$ the "hidden particle field" inside a particle while $\chi$ the "hidden antiparticle field" (rather than the "negative-energy component") inside the same particle.

Let us try to reinterpret the phenomena displayed in the kinematics of special relativity (SR) via the enhancement of $\chi$ field in a particle \cite{19,20}:

(a) Lorentz transformation

Consider a particle's WF shown by Eq.(4.3a) in an inertial frame $S$ (laboratory). Then take another $S'$ frame resting on the particle, so $p'=0$ and $E'=E_0=mc^2$. The WF in $S'$ frame reads:
\begin{equation}\label{4-3}
\psi(x',t')\sim \exp[\frac{i}{\hbar}(p'x'-E't')]=\exp[-\frac{i}{\hbar}E_0t']
\end{equation}
Here the space-time coordinates ($x',t'$) are introduced and defined in the $S'$ frame via the phase of WF as follows: Based on the assertion that "phase remains invariant under the coordinate transformation" which was named the "law of phase harmony " by de Broglie and was regarded by himself as the fundamental achievement all his life \cite{28}, comparing the phase in Eq.(4.4) with that in Eq.(4.3a) and using $E=E_0/\sqrt{1-v^2/c^2},p=Ev/c^2$, one finds
\begin{equation}\label{4-4}
t'=\dfrac{t-vx/c^2}{\sqrt{1-v^2/c^2}}
\end{equation}
Then, all formulas in the Lorentz transformation can be obtained. In some sense, what used here is a particle's wave-packet which serves as a microscopic "ruler", also a "clock" simultaneously.

(b) There is a speed limit c for a massive particle.

For a free KG particle, using Eq.(3.33), we may define an "impurity ratio" $R$ for the amplitude of hidden $\chi$ field to that of $\phi$ field and calculate it being
\begin{equation}\label{4-5}
R_{free}^{KG}=\dfrac{|\chi|}{|\phi|}=\left[\dfrac{1-\sqrt{1-(v/c)^2}}{1+\sqrt{1-(v/c)^2}}\right]
\end{equation}
When $v\to 0,\,|\chi|\to0$, with the increase of $v$, $|\chi|/|\phi|$ increases monotonously. The particle becomes more and more "impure" until $|\chi|/|\phi|\to 1$ as a limit of particle being still a particle. As shown by Eq.(4.6), the reason why its velocity has a limiting value $c$ (the speed of light) is because $\phi$ and $\chi$ have opposite evolution tendencies in space-time as shown by Eqs.(4.1)-(4.3) essentially, $\chi$ strives to hold $\phi$ back from going forward until a balance nearly reached when $|\chi|\to|\phi|$ and $v\to c$.

(c) The "length contraction" (FitzGerald-Lorentz contraction) and "time dilation"

As usual, we will show "length contraction" via a wave-packet of KG particle moving at a high-speed ($v$) but further ascribe it to the enhancement of $\chi$ field hidden inside the particle.

First, consider a wave-packet of KG particle at rest \cite{20,29}
\begin{equation}\label{4-6}
\psi(x,t)=(4\sigma\pi^3)^{-1/4}\int_{-\infty}^\infty\exp(-\frac{k^2}{2\sigma})\exp[i(kx-\omega t)]dk
\end{equation}
Assuming $\sqrt{\sigma}\ll\frac{mc}{\hbar}$, we have approximately that
\begin{equation}\label{4-7}
\psi(x,t)\cong\dfrac{(\sigma/\pi)^{1/4}}{(1+i\sigma\hbar t/m)^{1/2}}\exp\left[-\dfrac{\sigma x^2}{2(1+i\sigma\hbar t/m)}-\dfrac{imc^2t}{\hbar}\right]
\end{equation}
If $\sigma\hbar t/m\ll1$, the diffusion of wave-packet at low speed ($v\ll c$) can be ignored. Then we perform a "boost transformation" ($x\to (x-vt)/\sqrt{1-\beta^2},t\to(t-vx/c^2)/\sqrt{1-\beta^2},\beta=v/c$) to push the wave-packet to high velocity ($v\to c$), yielding
\begin{equation}\label{4-8}
\psi_{boost}(x,t)=(\frac{\sigma}{\pi})^{1/4}e^{i\alpha\xi}\exp(-i\frac{mc^2}{\hbar}\sqrt{1-\beta^2}t)\exp(-\frac{\xi^2}{2\varpi^2})
\end{equation}
where $\xi=\frac{mc}{\hbar}(x-vt),\alpha=\beta/\sqrt{1-\beta^2}$ and
\begin{equation}\label{4-9}
\varpi=\dfrac{mc\sqrt{1-\beta^2}}{\hbar\sqrt\sigma}\propto \sqrt{1-\beta^2}
\end{equation}
Here $\varpi$ is the width of wave-packet measured from its center $\xi=0$. Eqs.(4.7)-(4.10) show the "length contraction".

Second, we calculate from Eqs.(4.9) and (3.33) the values of $|\phi|^2,|\chi|^2$ and the probability density $\rho=|\phi|^2-|\chi|^2$ respectively. \footnotemark[1]\footnotetext[1]{Some pictures of numerical calculation are shown in Ref.\cite{29} and section 9.5C at Ref.\cite{20}, where an error in Eq.(9.5.26) is corrected here.} Their peak values all increase with the increase of $v$ (boost effect). However, the "intensity" of $|\phi|^2$ or $|\chi|^2$ increases even faster than that of $\rho$ while keeping the constraint $|\phi|>|\chi|$ in the boosting process.

We also calculate the square of "impurity ratio" $R$ for this moving wave-packet:
\begin{equation}\label{4-10}
[R_{free}^{KG}]^2=\dfrac{\int_{-\infty}^\infty|\chi|^2dx}{\int_{-\infty}^\infty|\phi|^2dx}=\left[\dfrac{1-\sqrt{1-(v/c)^2}}{1+\sqrt{1-(v/c)^2}}\right]^2
\end{equation}
which is the counterpart of Eq.(4.6) for a plane WF of KG particle.

With these calculations, we might intuitively understand the length contraction as an effect of coupling (\ie~ entanglement) between $\phi$ and $\chi$ fields due to their opposite evolution tendencies in space as discussed in previous point (b).

Let's turn to the "time dilation" shown by the variation of the mean life
\begin{equation}\label{4-11}
\tau=\dfrac{\tau_0}{\sqrt{1-\beta^2}}
\end{equation}
of a particle, say, a pion ($\pi^-$ or $\pi^+$) with its velocity $v$.

To understand it, let's return back to Eqs.(4.1)-(4.3) at $x=0$ and view the WF $\psi(\psi_c)$ on its complex phase with $Re\psi$ and $Im\psi$ ($Re\psi_c$ and $Im\psi_c$) as abscissa and ordinate. We may see that the time reading of the "inner clock" for a particle (or an antiparticle) is "clockwise" (or "counter clockwise"). Thus with the increase of particle velocity, though the time reading remains clockwise (due to the dominance of $\phi$ field), it runs slower and slower because of the enhancement of hidden $\chi$ field.

(d) WF's group velocity $u_g$ versus phase velocity $u_p$.

In RQM, a particle's velocity $v$ should be identified with its group velocity $u_g$. Actually, we have
\begin{equation}\label{4-12}
u_g=\dfrac{d\omega}{dk}=\dfrac{dE}{dp}=\dfrac{d}{dp}\sqrt{p^2c^2+m^2c^4}=\dfrac{pc^2}{E}=v\xrightarrow[E\to\infty]{} c
\end{equation}
However, the fact that there is an upper bound for particle's velocity doesn't mean that no speed can exceed that of light, $c$.
Indeed, there is another velocity $u_p$, the phase velocity in the WF
\begin{equation}\label{4-13}
u_p=\dfrac{\omega}{k}=\dfrac{E}{p}
\end{equation}
And the relation $E^2=p^2c^2+m^2c^4$ implies that\footnotemark[1]\footnotetext[1]{In 1923, de Broglie discovered Eq.(4.15) in his relativistic theory. However, in the Schr\"{o}dinger equation of NRQM, the phase velocity remains undefined. See Ref.[28].}
\begin{equation}\label{4-14}
u_gu_p=c^2,\quad u_p=\frac{c^2}{u_g}=\frac{c^2}{v}
\end{equation}
In our opinion, the role of $u_p>c$ here is crucial to maintain the quantum coherence of WF in the space-time globally, we will further discuss this problem elsewhere.

\section{V. Dirac Equation as Coupled equations of two-component Spinors}
\label{sec:DiracEquation}
\setcounter{equation}{0}
\renewcommand{\theequation}{5.\arabic{equation}}

Let us turn to the Dirac equation describing an electron
\begin{equation}\label{66}
\left(i\hbar\dfrac{\partial}{\partial t}-V\right)\psi=H\psi=(-i\hbar c{\boldsymbol\alpha}\cdot\nabla+\beta mc^2)\psi
\end{equation}
with ${\boldsymbol\alpha}$ and $\beta$ being $4\times4$ matrices, the WF $\psi$ is a four-component spinor
\begin{equation}\label{67}
\psi=\begin{pmatrix}\phi\\ \chi\end{pmatrix}
\end{equation}
Usually, the two-component spinors $\phi$ and $\chi$ are called "positive" and "negative" energy components. In our point of view, they are the hiding "particle" and "antiparticle" fields in a particle (electron) respectively (\cite{20}, see below). Substitution of Eq.(\ref{67}) into Eq.(\ref{66}) leads to
\begin{equation}\label{68}
\left\{
\begin{array}{l}
\left(i\hbar\dfrac{\partial}{\partial t}-V\right)\phi=-i\hbar c{\boldsymbol\sigma}\cdot\nabla\chi+mc^2\phi\\[3mm]
\left(i\hbar\dfrac{\partial}{\partial t}-V\right)\chi=-i\hbar c{\boldsymbol\sigma}\cdot\nabla\phi-mc^2\chi
\end{array}\right.
\end{equation}
(${\boldsymbol\sigma}$ are Pauli matrices). Eq.(\ref{68}) is invariant under the combined space-time inversion with
\begin{equation}\label{69}
\left\{
\begin{array}{l}
{\bf x}\to -{\bf x},t\to -t,\\
\phi({\bf x},t)\to {\cal C}\phi({\bf x},t)=\chi_c({\bf x},t),\;\chi({\bf x},t)\to {\cal C}\chi({\bf x},t)=\phi_c({\bf x},t)\\
V({\bf x},t)\to -V({\bf x},t)=V_c({\bf x},t)
\end{array}\right.
\end{equation}
showing that in its form of two-component spinors, Dirac equation is in conformity with the underlying symmetry Eq.(3.29).
Note that under the space-time inversion, the ${\boldsymbol\sigma}$ remain unchanged (However, see Eqs.(\ref{74})-(\ref{76}) below). Alternatively, Eq.(\ref{68}) also remains invariant under a mass inversion as
\begin{equation}\label{70}
m\to -m,\;\phi({\bf x},t)\to\chi_c({\bf x},t),\;\chi({\bf x},t)\to\phi_c({\bf x},t),\; V\to V
\end{equation}
In either case of Eq.(\ref{69}) or (\ref{70}), we have\footnotemark[2]\footnotetext[2]{The reason why we use $\psi'_c$ instead of $\psi_c$ will be clear in Eqs.(5.12)-(5.15). Actually, we emphasize Dirac equation as a coupling equation of two two-component spinors, Eq.(\ref{68}), rather than merely a four-component spinor equation.}
\begin{equation}\label{71}
\psi({\bf x},t)=\begin{pmatrix}\phi({\bf x},t)\\ \chi({\bf x},t)\end{pmatrix}\to \begin{pmatrix}\chi_c({\bf x},t)\\ \phi_c({\bf x},t)\end{pmatrix}=\psi'_c({\bf x},t)
\end{equation}
For concreteness, we consider a free electron moving along the $z$ axis with momentum $p=p_z>0$ and having a helicity $h={\boldsymbol\sigma}\cdot{\bf p}/|{\bf p}|=1$, its WF reads:
\begin{equation}\label{72}
\psi(z,t)\sim \begin{pmatrix}\phi\\ \chi\end{pmatrix}\sim
\begin{pmatrix}1\\ 0\\ \dfrac{p}{E+m}\\ 0\end{pmatrix}\exp[i(pz-Et)]
\end{equation}
with $|\phi|>|\chi|$. Under a space-time inversion ($z\to -z,t\to -t,p\to p_c,E\to E_c$) or mass inversion ($m\to -m,p\to -p_c,E\to -E_c$), it is transformed into a WF for positron (moving along $z$ axis)
\begin{equation}\label{73}
\psi'_c(z,t)\sim \begin{pmatrix}\chi_c\\ \phi_c\end{pmatrix}\sim
\begin{pmatrix}1\\ 0\\ \dfrac{p_c}{E_c+m}\\ 0\end{pmatrix}\exp[-i(p_cz-E_ct)]
\end{equation}
with $|\chi_c|>|\phi_c|,\;(p_c>0,E_c>0)$. However, the positron's helicity becomes $h_c=\frac{{\boldsymbol\sigma}_c\!\cdot\!{\bf p}_c}{|{\bf p}_c|}\!=\!-1$. This is because the total angular momentum operator for an electron reads
\begin{equation}\label{74}
\hat{\bf J}=\hat{\bf L}+\frac{\hbar}{2}{\boldsymbol\sigma}
\end{equation}
Under a space-time inversion, the orbital angular momentum operator is transformed as
\begin{equation}\label{75}
\hat{\bf L}={\bf r}\times\hat{\bf p}={\bf r}\times(-i\hbar\nabla)\to{-\bf r}\times(i\hbar\nabla)
=-{\bf r}\times\hat{\bf p}_c=-\hat{\bf L}_c
\end{equation}
To get $\hat{\bf J}\to -\hat{\bf J}_c$ with $\hat{\bf J}_c=\hat{L}_c+\frac{\hbar}{2}\hat{\boldsymbol\sigma}_c$, we should have
\begin{equation}\label{76}
\hat{\boldsymbol\sigma}_c=-\hat{\boldsymbol\sigma}
\end{equation}
Hence the values of matrix element for positron's spin operator ${\boldsymbol\sigma}_c$ is just the negative to that for ${\boldsymbol\sigma}$ in the same matrix representation.

Notice that Eq.(\ref{72}) describes an electron with positive helicity, \ie, ${\boldsymbol\Sigma}\cdot\hat{{\bf p}}\psi=p_z\psi=p\psi$ \footnotemark[2]\footnotetext[2]{${\boldsymbol\Sigma}=\begin{pmatrix}{\boldsymbol\sigma}&0\\ 0&{\boldsymbol\sigma}\end{pmatrix},\;{\boldsymbol\Sigma}_c=\begin{pmatrix}{\boldsymbol\sigma}_c&0\\ 0&{\boldsymbol\sigma}_c\end{pmatrix}$}. Under a space-time inversion, it is transformed into $(-{\boldsymbol\Sigma}_c)\cdot\hat{{\bf p}_c}\psi'_c=\Sigma_z(i\hbar\frac{\partial}{\partial z})\psi'_c=p_c\psi'_c$ in Eq.(\ref{73}), \ie, ${\boldsymbol\Sigma}_c\cdot\hat{{\bf p}_c}\psi'_c=-p_c\psi'_c$, meaning that Eq.(\ref{73}) describes a positron with negative helicity.

In its form of four-component spinor, Dirac equation, Eq.(5.1) with $V=0$, is usually written in a covariant form as (Pauli metric is used: $x_4=ict, \gamma_k=-i\beta\alpha_k, \gamma_4=\beta, \gamma_5=\gamma_1\gamma_2\gamma_3\gamma_4=-\begin{pmatrix}0 & I\\ I & 0\end{pmatrix}$, see Ref.\cite{24}):
\begin{equation}\label{17}
 (\gamma_\mu\partial_\mu+m)\psi=0
\end{equation}
Under a space-time (or mass) inversion, it turns into an equation for antiparticle:
\begin{equation}\label{78}
 (-\gamma_\mu\partial_\mu+m)\psi'_c=0
\end{equation}
with an example of $\psi'_c$ shown in Eq.(\ref{73}). Let us perform a representation transformation:
\begin{equation}\label{79}
\psi'_c\to\psi_c=(-\gamma_5)\psi'_c=\begin{pmatrix}\phi_c\\ \chi_c\end{pmatrix}
\end{equation}
and arrive at
\begin{equation}\label{80}
 (\gamma_\mu\partial_\mu+m)\psi_c=0
\end{equation}
due to $\{\gamma_5,\gamma_\mu\}=0$. Since $\psi_c$ and $\psi'_c$ are essentially the same in physics, (this is obviously seen from its resolved form, Eq.(\ref{68})), it is merely a trivial thing to change the position of $\chi_c$ in the 4-component spinor (lower in Eq.(\ref{79}) and upper in Eq.(\ref{73})). What important is $|\chi_c|>|\phi_c|$ for characterizing an  antiparticle versus $|\phi|>|\chi|$ for a particle. Therefore, if a particle with energy $E$ runs into a potential barrier $V=V_0>E+m$, its kinetic energy ($T=E-V_0<0$) becomes negative, and its WF's third component in Eq.(\ref{72}) suddenly turns into $\frac{p'}{E-V_0+m}=\frac{-p'}{V_0-E-m},(p'=\sqrt{(E-V_0)^2-m^2})$, whose absolute magnitude is larger than that of the first component. This means that it is an antiparticle's WF satisfying Eq.(\ref{80}) (with $E_c=V_0-E(>m)$ and  $|\chi_c|>|\phi_c|$) and will be crucial for the explanation of Klein paradox in Dirac equation (For detail, please see Appendix). However, we need to discuss the "probability density" $\rho$ and "probability current density" ${\bf j}$ for a Dirac particle versus $\rho_c$ and ${\bf j}_c$ for its antiparticle. Different from that in KG equation, now we have
\begin{equation}\label{81}
\rho=\psi^\dag\psi=\phi^\dag\phi+\chi^\dag\chi\to\rho_c=\psi_c^\dag\psi_c=\chi_c^\dag\chi_c+\phi_c^\dag\phi_c
\end{equation}
which is positive definite for either particle or antiparticle. On the other hand, we have
\begin{equation}\label{82}
{\bf j}=c\psi^\dag{\boldsymbol\alpha}\psi=c(\phi^\dag{\boldsymbol\sigma}\chi+\chi^\dag{\boldsymbol\sigma}\phi)
\to{\bf j}_c=c\psi_c^\dag{\boldsymbol\alpha}\psi_c=c(\chi_c^\dag{\boldsymbol\sigma}\phi_c+\phi_c^\dag{\boldsymbol\sigma}\chi_c)
\end{equation}
(we prefer to keep ${\boldsymbol\sigma}$ rather than ${\boldsymbol\sigma}_c$ for antiparticle). For Eqs.(\ref{72}), (\ref{73}) and (\ref{79}), we find ($c=\hbar=1$)
\begin{equation}\label{83}
j_z\sim\dfrac{2p}{E+m}>0\to j_z^c\sim\dfrac{2p_c}{E_c+m}>0\quad (V=0)
\end{equation}
which means that the probability current is always along the momentum's direction for either a particle or antiparticle.

Above discussions at RQM level may be summarized as follows: The first symptom for the appearance of an antiparticle is: If we perform an energy operator ( $E=i\hbar \partial / \partial t$) on a WF and find a negative energy ($E<0$) or a negative kinetic energy ($E-V <0$), we'd better to doubt the WF being a description of antiparticle and use the operators for antiparticle, Eq.(2.18). Then for further confirmation, two more criterions for $\rho$ and ${\bf j}$ are needed (see Appendix).

\section{VI. The Strong Reflection Invariance in CPT Theorem and QFT}
\label{sec:DiracEquation}
\setcounter{equation}{0}
\renewcommand{\theequation}{6.\arabic{equation}}

In QFT, the starting point is the field operator which is constructed for free complex boson field as \cite{30}
\begin{equation}\label{5-2}
\left\{
  \begin{array}{ll}
   \hat{\psi}({\bf x},t)=\sum\limits_{\bf p}\dfrac{1}{\sqrt{2V\omega_{\bf p}}}\left\{\hat{a}_{\bf p}\exp[i({\bf p}\cdot{\bf x}-Et)]
+\hat{b}^\dag_{\bf p}\exp[-i({\bf p}\cdot{\bf x}-Et)]\right\}\\[6mm]
   \hat{\psi}^\dag({\bf x},t)=\sum\limits_{\bf p}\dfrac{1}{\sqrt{2V\omega_{\bf p}}}\left\{\hat{a}^\dag_{\bf p}\exp[-i({\bf p}\cdot{\bf x}-Et)]
+\hat{b}_{\bf p}\exp[i({\bf p}\cdot{\bf x}-Et)]\right\}
  \end{array}
\right.
\end{equation}
Similarly, the field operator for free Dirac field reads:
\begin{equation}\label{5-10}
\left\{
  \begin{array}{ll}
   \hat{\psi}({\bf x},t)=\dfrac{1}{\sqrt{V}}\sum\limits_{\bf p}\sum\limits_{h=\pm1}\sqrt{\dfrac{m}{E}}\left[\hat{a}_{\bf p}^{(h)}u^{(h)}({\bf p})e^{i({\bf p}\cdot{\bf x}-Et)}+\hat{b}^{(h)\dag}_{\bf p}v^{(h)}({\bf p})e^{-i({\bf p}\cdot{\bf x}-Et)}\right]\\[6mm]
   \hat{\psi}^\dag({\bf x},t)=\dfrac{1}{\sqrt{V}}\sum\limits_{\bf p}\sum\limits_{h=\pm1}\sqrt{\dfrac{m}{E}}\left[\hat{a}_{\bf p}^{(h)\dag}u^{(h)\dag}({\bf p})e^{-i({\bf p}\cdot{\bf x}-Et)}+\hat{b}^{(h)}_{\bf p}v^{(h)\dag}({\bf p})e^{i({\bf p}\cdot{\bf x}-Et)}\right]
  \end{array}
\right.
\end{equation}
In Eq.(6.1), the annihilation operator $\hat{a}_{\bf p}$ for particle and the creation operator $\hat{b}^\dag_{\bf p}$ for antiparticle in Fock space are introduced. In Eq.(6.2), instead of index $s$ ($=\pm1/2$, the spin's projection along the fixed $z$ axis in space), the helicity $h$ is used. See Ref.\cite{31}.

Let us return back to the CPT theorem proved by L\"{u}ders and Pauli in 1954-1957 \cite{8,9}. The proof of CPT theorem contains a crucial step being the construction of so-called "strong reflection", consisting in a reflection of space and time about some arbitrarily chosen origin, \ie~ ${\bf r}\to -{\bf r},t\to -t$.

Pauli proposed and explained the strong reflection in Ref.\cite{9} as follows: When the space-time coordinates change their sign, every particle transforms into its antiparticle simultaneously. The physical sense of the strong reflection is the substitution of every emission (absorption) operator of a particle by the corresponding absorption (emission) operator of its antiparticle. And there is no need to reverse the sign of the electric charge when the sign of space-time coordinates is reversed.

What Pauli claimed, in our understanding, means that under the strong reflection for boson field, one has
\begin{equation}\label{5-4}
\left\{
  \begin{array}{l}
   {\bf x}\to -{\bf x},t\to -t,\\
   \hat{a}_{\bf p}\leftrightarrows \hat{b}^\dag_{\bf p}, \hat{a}^\dag_{\bf p}\leftrightarrows \hat{b}_{\bf p}
  \end{array}
\right.
\end{equation}
The mutual transformation, Eq.(6.3), in Fock space ensures the field operators, Eq.(6.1), invariant under the strong reflection in the sense of (see also \cite{21}):
\begin{equation}\label{6-4}
\begin{array}{l}
\hat{\psi}({\bf x},t)\to (\widehat{{\cal P}{\cal T}})\hat{\psi}({\bf x},t)(\widehat{{\cal P}{\cal T}})^{-1}=\hat{\psi}(-{\bf x},-t)=\hat{\psi}({\bf x},t)\\[3mm]
\hat{\psi}^\dag({\bf x},t)\to (\widehat{{\cal P}{\cal T}})\hat{\psi}^\dag({\bf x},t)(\widehat{{\cal P}{\cal T}})^{-1}=\hat{\psi}^\dag(-{\bf x},-t)=\hat{\psi}^\dag({\bf x},t)
\end{array}
\end{equation}
Here let us introduce the notation $\widehat{{\cal P}{\cal T}}$ to represent the strong reflection so that the presentation could be easier and clear as shown above. Similarly, for Dirac field, under the strong reflection one has
\begin{equation}\label{6-5}
\left\{
  \begin{array}{l}
   {\bf x}\to -{\bf x},t\to -t,\\
 \hat{a}^{(h)}_{\bf p}\leftrightarrows \hat{b}^{(-h)\dag}_{\bf p}, \hat{a}^{(h)\dag}_{\bf p}\leftrightarrows \hat{b}^{(-h)}_{\bf p}
\end{array}
\right.
\end{equation}
Here it is important to notice that the helicity, $h$, will be reversed before and after the strong reflection for a particle and its antiparticle respectively as discussed in section V. Because Eq.(6.2) is written in 4 component spinor covariant form, the invariance of Dirac field operator under the strong reflection should be expressed rigorously as
\begin{equation}\label{518}
\begin{array}{l}
\hat{\psi}({\bf x},t)\to (\widehat{{\cal P}{\cal T}})\hat{\psi}({\bf x},t)(\widehat{{\cal P}{\cal T}})^{-1}=-\gamma_5\hat{\psi}(-{\bf x},-t)=\hat{\psi}({\bf x},t)\\[3mm]
\hat{\psi}^\dag({\bf x},t)\to (\widehat{{\cal P}{\cal T}})\hat{\psi}^\dag({\bf x},t)(\widehat{{\cal P}{\cal T}})^{-1}=\hat{\psi}^\dag(-{\bf x},-t)(-\gamma_5)=\hat{\psi}^\dag({\bf x},t)
\end{array}
\end{equation}
\begin{equation}\label{519}
\hat{\psi}(-{\bf x},-t)=-\gamma_5\hat{\psi}({\bf x},t),\quad \hat{\psi}^\dag(-{\bf x},-t)=\hat{\psi}^\dag({\bf x},t)(-\gamma_5)
\end{equation}
which are useful in proving the "spin-statistics connection" by strong reflection invariance.

QFT is a successful theory just because it is established on sound basis with the field operator being one of its cornerstones. Historically, through various trials and checks, Eqs.(6.1)-(6.2) were eventually found (see section 3.5 of Ref.\cite{30}). Why they are correct and why one would fail otherwise? In our understanding, it is just because they are invariant under the strong reflection as shown by Eqs.(6.4) and (6.6).

However, as emphasized by Pauli \cite{9} and further stressed by L\"{u}ders \cite{8}, at least two more rules should be added in doing calculations:

(a) The order of an operator product in Fock space has to be reversed under the strong reflection, \eg, $(\widehat{{\cal P}{\cal T}})\hat{A}\hat{B}(\widehat{{\cal P}{\cal T}})^{-1}=(\widehat{{\cal P}{\cal T}})\hat{B}(\widehat{{\cal P}{\cal T}})^{-1}(\widehat{{\cal P}{\cal T}})\hat{A}(\widehat{{\cal P}{\cal T}})^{-1}$. So is the order of a process occurred in a many-particle system.

(b) Another rule is: One should always take the normal ordering when dealing with quadratic forms like $\hat{\bar\psi}(x)\hat{\psi}(x)$ \etc

Then Pauli and L\"{u}ders were able to prove that the Hamiltonian density ${\cal H}({\bf x},t)$ for a broad kind of model in relativistic QFT is invariant under an operation of "strong reflection", \ie,
\begin{equation}\label{5-23}
{\hat{\cal H}}({\bf x},t)\to\widehat{{\cal P}{\cal T}}{\hat{\cal H}}({\bf x},t)(\widehat{{\cal P}{\cal T}})^{-1}={\hat{\cal H}}(-{\bf x},-t)={\hat{\cal H}}({\bf x},t)
\end{equation}
The Hamiltonian density is also invariant under a Hermitian conjugation (H.C.) as:
\begin{equation}\label{5-24}
{\hat{\cal H}}({\bf x},t)\to{\hat{\cal H}}^\dag({\bf x},t)={\hat{\cal H}}({\bf x},t)
\end{equation}
Furthermore, they proved the CPT theorem via the identification of the product of T,C, and P in QFT with the combined operation of the strong reflection and a Hermitian conjugation.

The validity of CPT invariance, \ie~ Eqs.(\ref{5-23})-(\ref{5-24}) has been verified experimentally since the discovery of parity violation (\cite{3,4,5,6} \etc) and the establishment (and development) of standard model (\cite{32} \etc) in particle physics till this day. See the excellent book, Ref.\cite{15} and the Review of Particle Physics, Ref.\cite{7}.

After restudying the historical contribution of Pauli-L\"{u}ders strong reflection invariance, we feel good in understanding that what we claim in RQM (sections III-V) is essentially the same as or very close to their idea.

In fact, this paper is the direct continuation of our first one in 1974 \cite{18}, which was inspired jointly by the discoveries of violations in P, C, CP, T symmetries individually (but CPT invariance holds), also by Lee-Wu's proposal in 1965 that the relationship between a particle $|a\rangle$ and its antiparticle $|\bar{a}\rangle$ should be \cite{10}:
\begin{equation}\label{6-10}
|\bar{a}\rangle=CPT|a\rangle
\end{equation}
and especially by Pauli's invention of the strong reflection in 1955 \cite{9}.

Below, we would like to show that WFs for a particle and its antiparticle given in Eqs.(5.7)-(5.8) are precisely that derived from QFT as expected.

Using Eq.(6.2) for Dirac field, we find the WF of an electron being
\begin{equation}\label{6-11}
\psi_{e^-}({\bf x},t)=\langle0|\hat{\psi}({\bf x},t)|e^-,{\bf p}_1,h_1\rangle=\langle0|\hat{\psi}({\bf x},t)\hat{a}^{(h_1)\dag}_{\bf p_1}|0\rangle=\frac{1}{\sqrt V}\sqrt{\frac{m}{E_1}}u^{(h_1)}({\bf p}_1)e^{i({\bf p}_1\cdot{\bf x}-E_1t)}
\end{equation}
but the hermitian conjugate of a positron's WF is given by
\begin{equation}\label{6-12}
\psi^\dag_{e^+}({\bf x},t)=\langle0|\hat{\psi}^\dag({\bf x},t)|e^+,{\bf p}_c,h_c\rangle=\langle0|\hat{\psi}^\dag({\bf x},t)\hat{b}^{(h_c)\dag}_{\bf p_c}|0\rangle=\frac{1}{\sqrt V}\sqrt{\frac{m}{E_c}}v^{(h_c)\dag}({\bf p}_c)e^{i({\bf p}_c\cdot{\bf x}-E_ct)}
\end{equation}
which leads to positron's WF being
\begin{equation}\label{6-13}
\psi_{e^+}({\bf x},t)=\frac{1}{\sqrt V}\sqrt{\frac{m}{E_c}}v^{(h_c)}({\bf p}_c)e^{-i({\bf p}_c\cdot{\bf x}-E_ct)}
\end{equation}
Similarly, Eqs.(2.20)-(2.21) can be derived from Eq.(6.1) as expected.

\section{VII. An Oversight in QFT (Helicity States or Spin States?)--- Why a Parity-Violation Phenomenon Was Overlooked Since 1956-1957?}
\label{sec:discussion}
\setcounter{equation}{0}
\renewcommand{\theequation}{7.\arabic{equation}}

Through analysis in RQM till QFT, we stress the necessity of using helicity ($h$) to describe a fermion or antifermion. Here is an interesting example. Since 2002, Shi and Ni \cite{33,34,35,36} predicted a parity-violation phenomenon as follows:

An unstable (decaying) fermion (\eg, neutron or muon) has different mean lifetimes for being right-handed (RH) or left-handed (LH) polarized during its flight with the same speed $v\,(\beta=v/c)$
\begin{equation}\label{}
\tau_R=\dfrac{\tau}{1-\beta},\quad \tau_L=\dfrac{\tau}{1+\beta}
\end{equation}
where $\tau=\tau_0/\sqrt{1-\beta^2}$, $\tau_0$ the mean lifetime when it is at rest. Similarly, for its antifermion, their lifetimes will be
\begin{equation}\label{}
\bar{\tau}_R=\dfrac{\tau}{1+\beta},\quad \bar{\tau}_L=\dfrac{\tau}{1-\beta}
\end{equation}
Hence, the lifetime asymmetry can be defined as
\begin{equation}\label{}
A=\dfrac{\tau_R-\tau_L}{\tau_R+\tau_L}=\beta
\end{equation}
This is not a small effect. For instance, in Fermilab, physicists consider to build a muon collider  \cite{37}. The collision of $\mu^-$ and $\mu^+$ beams must happen before the muons decay. It was estimated that if a muon rings along at 1.5~TeV, the time dilation of SR stretches its lifetime to 30 milliseconds --- up from 2 microseconds when it's still. That's time enough for 500 circuits in the final ring. However, as discussed in Ref.\cite{36}, if the prediction of life asymmetry Eq.(7.1) is correct, the lifetime of RH $\mu^-$ will be stretched to 146 days while that of LH $\mu^-$ only 15 milliseconds. The lifetime asymmetry of $\mu^+$ will be just the opposite as shown by Eq.(7.2). Therefore, it seems necessary to take Eqs.(7.1)-(7.2) into account in the design of a muon collider.

The problem is: How can such a parity-violation phenomenon be overlooked since 1956-1957? One theoretical reason is: in the past, for describing a fermion in flight (${\bf v}\neq0$), instead of helicity states, the "spin-states" assigned by $s$ (spin's projection along the fixed $z$ axis in space) were often incorrectly used (see \cite{34,35}). So previous calculations on the lifetime always led to a prediction that $\tau=\tau_0/\sqrt{1-\beta^2}$ without parity-violation in contrast to Eqs.(7.1)-(7.3). \footnotemark[1]\footnotetext[1]{The wonderful experiment by Wu {\it et al}. \cite{4} reveals the decay configuration of a polarized neutron bearing a strong resemblance to a "comet" with its "head" oriented along neutron's spin parallel to $z$ axis in space (note that a static neutron has no helicity $h$, see \cite{38}) while its "tail" composed of emitted $e^-$ and $\bar{\nu}_e$. So it was expected intuitively that \cite{33} if one pushes the "comet" along its "head"'s direction, it (suddenly has a helicity $h=1$ and) will be relatively more stable than it is pushed along its "tail" (when it has $h=-1$). That's what Eq.(7.1) means and why the use of "spin state" fails to get it right. }

The interesting thing is: While Eqs.(7.1) and (7.2) display the violation of P or C symmetry to its maximum, their "cross-symmetry", $\tau_R=\bar{\tau}_L$ and $\tau_L=\bar{\tau}_R$, reflects the symmetry of ${\cal P}{\cal T}={\cal C}$ shown by Eq.(6.5) exactly.

\section{VIII. Dirac Particles Conserve the Parity Whereas Neutrinos are Likely the Tachyons}
\label{sec:discussion}
\setcounter{equation}{0}
\renewcommand{\theequation}{8.\arabic{equation}}

\subsection{VIIIA. Why Dirac Equation Respects the Parity Symmetry?}

In the standard representation of Dirac equation for free particle ($\hbar=c=1$)
\begin{equation}
i\dfrac{\partial}{\partial t}\psi^{(D)}=-i{\boldsymbol\alpha}\cdot\nabla\psi^{(D)}+\beta m\psi^{(D)}
\end{equation}
Let us choose ${\boldsymbol\alpha}=-\begin{pmatrix}0 & {\boldsymbol\sigma}\\{\boldsymbol\sigma}& 0\end{pmatrix},\beta=\begin{pmatrix}I & 0\\0& -I\end{pmatrix},\psi^{(D)}=\begin{pmatrix}\phi^{(D)}\\\chi^{(D)} \end{pmatrix}$, then
\begin{equation}
 \left\{
   \begin{array}{l}
     i\dfrac{\partial}{\partial t}\phi^{(D)}=i{\boldsymbol\sigma}\cdot\nabla\chi^{(D)}+ m\phi^{(D)}\\[5mm]
     i\dfrac{\partial}{\partial t}\chi^{(D)}=i{\boldsymbol\sigma}\cdot\nabla\phi^{(D)}- m\chi^{(D)}
   \end{array}
 \right.
\end{equation}
As discussed in section V, Eqs.(8.1)-(8.2) are invariant under the space-time inversion:
\begin{equation}
 \left\{
   \begin{array}{l}
     {\bf x}\to -{\bf x},t\to -t\\[2mm]
     \phi^{(D)}({\bf x},t)\to \phi^{(D)}(-{\bf x},-t)=\chi_c^{(D)}({\bf x},t)\\[3mm]
     \chi^{(D)}({\bf x},t)\to \chi^{(D)}(-{\bf x},-t)=\phi_c^{(D)}({\bf x},t)
   \end{array}
 \right.
\end{equation}
with subscript "c" meaning the antiparticle.

After transforming $\psi^{(D)}$ into the "Weyl representation" (chiral representation) as
\begin{equation}
 \psi^{(D)}\to \dfrac{1}{\sqrt2}\begin{pmatrix}I & I\\I& -I\end{pmatrix}\begin{pmatrix}\phi^{(D)}\\\chi^{(D)} \end{pmatrix}=\begin{pmatrix}\xi^{(D)}\\ \eta^{(D)} \end{pmatrix}
\end{equation}
we have
\begin{equation}
 \left\{
   \begin{array}{l}
     i\dfrac{\partial}{\partial t}\xi^{(D)}=i{\boldsymbol\sigma}\cdot\nabla\xi^{(D)}+ m\eta^{(D)}\\[5mm]
     i\dfrac{\partial}{\partial t}\eta^{(D)}=-i{\boldsymbol\sigma}\cdot\nabla\eta^{(D)}+ m\xi^{(D)}
   \end{array}
 \right.
\end{equation}
If $m=0$, Eq.(8.5) reduces into two Weyl equations describing two kinds of permanently LH and RH polarized massless fermions respectively. So we may name $\xi^{(D)}$ and $\eta^{(D)}$ (which are usually called as chirality states or chiral fields in 4-component covariant form) as the "hidden LH and RH spinning fields" inside a Dirac particle, which can be either LH or RH polarized (with helicity $h=-1$ or 1) explicitly. See below.

A new symmetry is hidden in Eq.(8.5), which remains invariant under the pure space inversion (${\bf x}\to -{\bf x},t\to t$) transformation, \ie, the parity operation as
\begin{equation}
 \left\{
   \begin{array}{l}
     \xi^{(D)}({\bf x},t)\to \xi^{(D)}(-{\bf x},t)=\eta^{(D)'}({\bf x},t)\\
     \eta^{(D)}({\bf x},t)\to \eta^{(D)}(-{\bf x},t)=\xi^{(D)'}({\bf x},t)
   \end{array}
 \right.
\end{equation}
Here we add "$'$" in the superscript of RHS to stress that the WF after the space inversion may be different from that at the LHS (before the space inversion). We knew that the WF in Dirac representation after a space inversion reads
\begin{equation}
 \hat{P}\psi^{(D)}({\bf x},t)=\gamma_4\psi^{(D)}(-{\bf x},t)
\end{equation}
Using Eq.(8.6), the RHS of Eq.(8.7) turns out to be
\begin{equation}
\dfrac{1}{\sqrt2}\gamma_4\begin{pmatrix}\xi^{(D)}(-{\bf x},t)+\eta^{(D)}(-{\bf x},t)\\\xi^{(D)}(-{\bf x},t)-\eta^{(D)}(-{\bf x},t)\end{pmatrix}=\dfrac{1}{\sqrt2}\begin{pmatrix}\xi^{(D)'}({\bf x},t)+\eta^{(D)'}({\bf x},t)\\\xi^{(D)'}({\bf x},t)-\eta^{(D)'}({\bf x},t)\end{pmatrix}=\psi^{(D)'}({\bf x},t)
\end{equation}
Hence, we understand the reason why a Dirac particle respects the parity symmetry as shown by Eq.(8.7) is because it enjoys the symmetry, Eq.(8.6) hiding in the 2-component spinor form (in Weyl representation).

For concreteness, let's write down the solution of Eq.(8.1)
\begin{equation}\label{}
\psi^{(D)}({\bf x},t)=\begin{pmatrix}\phi^{(D)}\\\chi^{(D)} \end{pmatrix}\sim\begin{pmatrix}\phi_0\\\dfrac{-{\boldsymbol\sigma}\cdot{\bf p}}{E+m}\phi_0 \end{pmatrix},\quad (E=\sqrt{{\bf p}^2+m^2}>0)
\end{equation}
Furthermore, we choose a simplest "spin state" with $\hat{\bf p}\psi^{(D)}=p_z\psi^{(D)}$ and $\hat{\sigma}_z\psi^{(D)}=\psi^{(D)}$:
\begin{equation}\label{}
\psi_{s_z=1/2}^{(D)}(z,t)=\begin{pmatrix}\phi^{(D)}\\\chi^{(D)} \end{pmatrix}\sim\begin{pmatrix}1\\0\\\dfrac{-p_z}{E+m}\\0\end{pmatrix}e^{i(p_zz-Et)}\,(E>0)
\end{equation}
while Eq.(8.10) is an eigenfunction of $\hat{\sigma}_z$ with eigenvalue $s_z=1/2$, its helicity $h$ remains unfixed, depending on the value of $p_z$ being positive or negative. Only after $p_z=p>0$ is fixed, can we have a "helicity state" describing a RH particle with $h=1$:
\begin{equation}\label{}
\psi_{RH}^{(D)}(z,t)=\begin{pmatrix}\phi^{(D)}\\\chi^{(D)} \end{pmatrix}\sim\begin{pmatrix}1\\0\\ \dfrac{-p}{E+m}\\0\end{pmatrix}e^{i(pz-Et)}\,(p>0,E>0)
\end{equation}
Looking at Eq.(8.11) in the Weyl representation, we see that
\begin{equation}\label{}
\xi^{(D)}=\dfrac{1}{\sqrt2}(\phi^{(D)}+\chi^{(D)})\sim\dfrac{1}{\sqrt2}\begin{pmatrix}1-\dfrac{p}{E+m}\\0\end{pmatrix},
\eta^{(D)}=\dfrac{1}{\sqrt2}(\phi^{(D)}-\chi^{(D)})\sim\dfrac{1}{\sqrt2}\begin{pmatrix}1+\dfrac{p}{E+m}\\0\end{pmatrix}
\end{equation}
$|\xi^{(D)}|<|\eta^{(D)}|$. So Eq.(8.11) describes a RH particle just because the $\eta^{(D)}$ field dominates the $\xi^{(D)}$ field. Now we perform a space inversion on Eq.(8.11), according to the rule Eq.(8.7), yielding
\begin{equation}\label{}
\begin{array}{l}
\hat{P}\psi_{RH}^{(D)}(z,t)\sim\begin{pmatrix}1\\0\\\dfrac{p}{E+m}\\0\end{pmatrix}e^{i(-pz-Et)}=\begin{pmatrix}\phi^{(D)'}\\\chi^{(D)'} \end{pmatrix}=\psi^{(D)'}(z,t)\\[4mm]
\xi^{(D)'}=\dfrac{1}{\sqrt2}(\phi^{(D)'}+\chi^{(D)'})\sim\dfrac{1}{\sqrt2}\begin{pmatrix}1+\dfrac{p}{E+m}\\0\end{pmatrix},
\eta^{(D)'}=\dfrac{1}{\sqrt2}(\phi^{(D)'}-\chi^{(D)'})\sim\dfrac{1}{\sqrt2}\begin{pmatrix}1-\dfrac{p}{E+m}\\0\end{pmatrix}\\[5mm]
\xi^{(D)'}=\dfrac{E+p}{m}\eta^{(D)'},\,|\xi^{(D)'}|>|\eta^{(D)'}|
\end{array}
\end{equation}
Hence we see that the reason why $\psi^{(D)'}(z,t)$ becomes a LH WF, \ie,
\begin{equation}
 \hat{P}\psi^{(D)}_{RH}(z,t)=\psi^{(D)}_{LH}(z,t)
\end{equation}
is just because of the dominance of $\xi^{(D)'}$ field over $\eta^{(D)'}$ field after the P-operation. Before and after the operation, $p\to -p$, the dominant (subordinate) field is transformed into dominant (subordinate) field: $\eta^{(D)}\to\xi^{(D)'}$, ($\xi^{(D)}\to\eta^{(D)'}$), as shown by Eq.(8.6).

In summary, Dirac equation is invariant under a space inversion whereas its concrete solution of WF may be not. The latter may change from that for a RH particle to a LH one or vice versa, but with the same mass $m$, showing the law of parity conservation exactly.

\subsection{VIIIB. Tachyon Equation as a Counterpart of the Dirac Equation}

Now a question arises: Can we find an equation which violates the symmetry of pure space inversion?

The answer is "yes". Let's introduce a new equation in Weyl representation from Eq.(8.5) by erasing the superscript (D), replacing the mass term by $m\to m_s$ and changing its sign from "+" to "-" in the first equation of Eq.(8.5) only
\begin{equation}
 \left\{
   \begin{array}{l}
     i\dfrac{\partial}{\partial t}\xi=i{\boldsymbol\sigma}\cdot\nabla\xi- m_s\eta\\[5mm]
     i\dfrac{\partial}{\partial t}\eta=-i{\boldsymbol\sigma}\cdot\nabla\eta+ m_s\xi
   \end{array}
 \right.
\end{equation}
where $m_s$ (real and positive) refers to the mass of a hypothetical particle. We will see immediately that it is a "superluminal particle" or "tachyon".

Indeed, substituting a plane-wave solution
\begin{equation}
\xi\sim\eta\sim\exp[i(p_zz-Et)]\begin{pmatrix}0\\ 1\end{pmatrix}
\end{equation}
with the particle's helicity $h=-1$ into Eq.(8.15), we find that ($p_z=p>0,E>0$)
\begin{equation}
  E^2 = p^2-m^2_s
\end{equation}
\begin{equation}
  \xi = \frac{1}{m_s}(p+E)\eta,\qquad |\xi|>|\eta|
\end{equation}
Since $E=\hbar\omega$ and ${\bf p}=\hbar{\bf k}$, from Eq.(8.17), the dispersion-relation of wave reads
\begin{equation}
\omega^2=k^2-m^2_s
\end{equation}
As in section IV, we define the wave's phase velocity $u_p$ as
\begin{equation}
u_p=\dfrac{\omega}{k}
\end{equation}
while its group velocity $u_g$
\begin{equation}
u_g=\dfrac{d\omega}{dk}=v
\end{equation}
being identical with the particle's velocity $v$. Eq.(8.19) yields a relation between them coinciding with Eq.(4.15) exactly:
\begin{equation}
u_pu_g=c^2
\end{equation}
However, the relations among $E,p$ and $v$ are dramatically different
\begin{equation}
E=\dfrac{m_sc^2}{\sqrt{\frac{v^2}{c^2}-1}},\quad
p=\dfrac{m_sv}{\sqrt{\frac{v^2}{c^2}-1}}
\end{equation}
which dictate $v>c$ such that $E,{\bf p}$ are real and $E>0$.

Like Eq.(8.4), we define:
\begin{equation}
\phi=\dfrac{1}{\sqrt2}(\xi+\eta),\quad
\chi=\dfrac{1}{\sqrt2}(\xi-\eta)
\end{equation}
and find from Eq.(8.15) that (in Dirac representation)
\begin{equation}\label{}
i\dfrac{\partial}{\partial t}\psi=-i{\boldsymbol\alpha}\cdot\nabla\psi+ \beta_sm_s\psi
\end{equation}
\begin{equation}
 \left\{
   \begin{array}{l}
     i\dfrac{\partial}{\partial t}\phi=i{\boldsymbol\sigma}\cdot\nabla\chi+ m_s\chi\\[5mm]
     i\dfrac{\partial}{\partial t}\chi=i{\boldsymbol\sigma}\cdot\nabla\phi- m_s\phi
   \end{array}
 \right.
\end{equation}
($\psi=\begin{pmatrix}\phi\\ \chi\end{pmatrix},\beta_s=\begin{pmatrix}0& I\\ -I & 0\end{pmatrix}$) Despite the difference between Eq.(8.26) and Dirac equation, Eq.(8.2), both of them respect the combined space-time inversion (${\cal PT}$) symmetry like Eq.(8.3)
\begin{equation}
 \left\{
   \begin{array}{l}
     \phi({\bf x},t)\to \phi(-{\bf x},-t)=\chi_c({\bf x},t)\\
     \chi({\bf x},t)\to \chi(-{\bf x},-t)=\phi_c({\bf x},t)
   \end{array}
 \right.
\end{equation}
with
\begin{equation}
 \left\{
   \begin{array}{l}
     i\dfrac{\partial}{\partial t}\chi_c=i{\boldsymbol\sigma}\cdot\nabla\phi_c-m_s\phi_c\\[4mm]
     i\dfrac{\partial}{\partial t}\phi_c=i{\boldsymbol\sigma}\cdot\nabla\chi_c+m_s\chi_c
   \end{array}
 \right.
\end{equation}
Similarly, we define the WF in Weyl representation after ${\cal PT}$ inversion as:
\begin{equation}
\left\{
   \begin{array}{l}
     \xi({\bf x},t)\to{\cal PT}\xi({\bf x},t)=\xi(-{\bf x},-t)=\eta_c({\bf x},t)\\
     \eta({\bf x},t)\to{\cal PT}\eta({\bf x},t)=\eta(-{\bf x},-t)=\xi_c({\bf x},t)
   \end{array}
 \right.
\end{equation}
Based on Eqs.(8.27)-(8.29), we find
\begin{equation}
\left\{
   \begin{array}{l}
     \eta_c({\bf x},t)=\dfrac{1}{\sqrt2}[\chi_c({\bf x},t)+\phi_c({\bf x},t)]\\[4mm]
     \xi_c({\bf x},t)=\dfrac{1}{\sqrt2}[\chi_c({\bf x},t)-\phi_c({\bf x},t)]
   \end{array}
 \right.
\end{equation}
\begin{equation}
 \left\{
   \begin{array}{l}
     i\dfrac{\partial}{\partial t}\eta_c=i{\boldsymbol\sigma}\cdot\nabla\eta_c+m_s\xi_c\\[4mm]
     i\dfrac{\partial}{\partial t}\xi_c=-i{\boldsymbol\sigma}\cdot\nabla\xi_c-m_s\eta_c
   \end{array}
 \right.
\end{equation}
which can also be obtained via the ${\cal PT}$ operation on Eq.(8.15). Eqs.(8.15) and (8.31) are better to be compared in the following form:
\begin{equation}
\left\{
   \begin{array}{l}
     \hat{E}\xi=-{\boldsymbol\sigma}\cdot\hat{\bf p}\xi-m_s\eta\\
     \hat{E}\eta={\boldsymbol\sigma}\cdot\hat{\bf p}\eta+m_s\xi
   \end{array}
 \right.
\end{equation}
\begin{equation}
\left\{
   \begin{array}{l}
     \hat{E}_c\eta_c={\boldsymbol\sigma}_c\cdot\hat{\bf p}_c\eta_c-m_s\xi_c\\
     \hat{E}_c\xi_c=-{\boldsymbol\sigma}_c\cdot\hat{\bf p}_c\xi_c+m_s\eta_c
   \end{array}
 \right.
\end{equation}
($\hat{E}_c=-i\frac{\partial}{\partial t},\hat{\bf p}_c=i\nabla,{\boldsymbol\sigma}_c=-{\boldsymbol\sigma}$). Interestingly, Eq.(8.33) can also be reached from Eq.(8.32) via a "mass inversion" like that in section III and V:
\begin{equation}
\left\{
   \begin{array}{l}
    m_s\to-m_s,\\
     \hat{E}\to -\hat{E}_c=i\frac{\partial}{\partial t}\,\qquad(\text{no change in $t$})\\
     \hat{\bf p}\to -\hat{\bf p}_c=-i\nabla\,\qquad(\text{no change in ${\bf x}$})\\
     {\boldsymbol\sigma}\to -{\boldsymbol\sigma}_c={\boldsymbol\sigma}\,\qquad(\text{no change in ${\boldsymbol\sigma}$})\\
    \xi({\bf x},t)\to \eta_c({\bf x},t),\,\eta({\bf x},t)\to \xi_c({\bf x},t)
   \end{array}
 \right.
\end{equation}
Furthermore, the probability density and probability current density before and after the ${\cal PT}$ inversion can be derived as:
\begin{equation}
   \begin{array}{l}
     \rho=\phi^\dag\chi+\chi^\dag\phi=\xi^\dag\xi-\eta^\dag\eta\\[2mm]
     \xrightarrow[{\cal PT}]{} \rho_c=\chi_c^\dag\phi_c+\phi_c^\dag\chi_c=\eta_c^\dag\eta_c-\xi_c^\dag\xi_c
   \end{array}
\end{equation}
and
\begin{equation}
   \begin{array}{l}
     {\bf j}=-(\phi^\dag{\boldsymbol\sigma}\phi+\chi^\dag{\boldsymbol\sigma}\chi)=-(\xi^\dag{\boldsymbol\sigma}\xi+\eta^\dag{\boldsymbol\sigma}\eta)\\[2mm]
     \xrightarrow[{\cal PT}]{} {\bf j}_c=-(\chi_c^\dag{\boldsymbol\sigma}\chi_c+\phi_c^\dag{\boldsymbol\sigma}\phi_c)=-(\eta_c^\dag{\boldsymbol\sigma}\eta_c+\xi_c^\dag{\boldsymbol\sigma}\xi_c)
   \end{array}
\end{equation}
respectively. It is the sharp contrast between Eq.(8.35) and Eq.(5.16) for Dirac equation (\ie, $\rho^{(D)}=\xi^{(D)\dag}\xi^{(D)}+\eta^{(D)\dag}\eta^{(D)}$), that makes Eq.(8.15) so unique as shown below.

Let us look at the example of WF for tachyon, Eqs.(8.16)-(8.18), with $E>0,p_z>0$ and $h=-1$. It is allowed just because $|\xi|>|\eta|$ and so $\rho>0$. Second choice of Eq.(8.16) with $p_z=-p<0\,(p>0),h=+1$ but
\begin{equation}\label{}
\xi=\frac{1}{m_s}(-p+E)\eta,\quad |\xi|<|\eta|
\end{equation}
should be fobidden due to its $\rho<0$. Another two possible WFs with $\xi\sim\eta\sim \begin{pmatrix}1\\0\end{pmatrix}$ have $p_z=p$ and $p_z=-p$ respectively, only the last one with $p_z=-p<0\,(p>0),h=-1$ is allowed due to its $|\xi|>|\eta|$ and $\rho>0$.

Let us turn to the solution of Eq.(8.31) for antitachyon with $E_c>0$ by just performing ${\cal PT}$ operation on Eq.(8.16) yielding:
\begin{equation}\label{}
\eta_c\sim\xi_c\sim \exp[-i(p_z^cz-E_ct)]\begin{pmatrix}0\\1\end{pmatrix}
\end{equation}
Now if $p_z^c=p_c(=|{\bf p}_c|)>0$, since $\sigma_z^c=-\sigma_z,\,\sigma_z^c\begin{pmatrix}0\\1\end{pmatrix}=1$, so helicity $h_c=1$. Substitution of Eq.(8.38) into Eq.(8.33) yields:
\begin{equation}\label{}
\eta_c=\frac{1}{m_s}(p_c+E_c)\xi_c,\quad |\eta_c|>|\xi_c|
\end{equation}
which is allowed due to $\rho_c>0$. Second choice of Eq.(8.38) with $p_z^c=-p_c<0\,(p_c>0),h=-1$ but
\begin{equation}\label{}
\eta_c=\frac{1}{m_s}(-p_c+E_c)\xi_c,\quad |\eta_c|<|\xi_c|
\end{equation}
should be forbidden due to its $\rho_c<0$. In another two possible WFs with $\eta_c\sim\xi_c\sim \begin{pmatrix}1\\0\end{pmatrix}$, only that with $p_z^c=-p_c<0,h_c=+1$ is allowed due to $\rho_c>0$.

Hence we see that: The tachyon can only exist in a left-handed (LH) polarized state (with helicity $h=-1$) whereas antitachyon only in a right-handed (RH) polarized state (with $h_c=1$). We tentatively link this strange feature with that found in neutrinos --- only $\nu_L$ and $\bar{\nu}_R$ exists in nature whereas $\nu_R$ and $\bar{\nu}_L$ are strictly forbidden.

Furthermore, at first sight, although Eq.(8.15) certainly has no symmetry under the space inversion (${\bf x}\to -{\bf x},t\to t$), it seems to enjoy a pure "time-inversion" (${\bf x}\to{\bf x},t\to -t$) symmetry like
\begin{equation}
 \left\{
   \begin{array}{l}
     \xi({\bf x},t)\to \xi({\bf x},-t)=\eta'_c({\bf x},t)\\
     \eta({\bf x},t)\to \eta({\bf x},-t)=\xi'_c({\bf x},t)
   \end{array}
 \right.
\end{equation}
\begin{equation}
 \left\{
   \begin{array}{l}
     i\dfrac{\partial}{\partial t}\eta'_c=-i{\boldsymbol\sigma}\cdot\nabla\eta'_c+m_s\xi'_c\\[4mm]
     i\dfrac{\partial}{\partial t}\xi'_c=i{\boldsymbol\sigma}\cdot\nabla\xi'_c-m_s\eta'_c
   \end{array}
 \right.
\end{equation}
We add "$'$" in the superscript of $\eta'_c$ to stress that $\eta'_c({\bf x},t)$ (being a time reversed WF), though looks like some antitachyon's WF, is obviously different from $\eta_c({\bf x},t)$ gained through the ${\cal PT}$ inversion, Eq.(8.31). Actually, based on Eqs.(8.29)-(8.31) and (8.41)-(8.42), we have:
\begin{equation}
\left\{
   \begin{array}{l}
\eta'_c({\bf x},t)=\eta_c(-{\bf x},t),\quad \xi'_c({\bf x},t)=\xi_c(-{\bf x},t)\\[3mm]
\eta'_c(-{\bf x},t)=\eta_c({\bf x},t),\quad \xi'_c(-{\bf x},t)=\xi_c({\bf x},t)
\end{array}
 \right.
\end{equation}
Interestingly, we cannot find from Eq.(8.42) the "physical solution" of $\eta'_c({\bf x},t)$ with $|\eta'_c|>|\xi'_c|$ (so $\rho_c>0$) and $h_c=1$ (for $\bar{\nu}_R$) simultaneously. Only $\eta'_c(-{\bf x},t)$ makes physical sense, but it is just $\eta_c({\bf x},t)$ like that discussed in Eq.(8.39). Notice that the sign change ${\bf x}\to -{\bf x}$ in the phase of WF makes a change in the direction of momentum ${\bf p}_c\to -{\bf p}_c$. But a WF is always composed of two fields in confrontation, like $\eta_c$ versus $\xi_c$ here. And the explicit helicity $h_c$ is determined by which one of these two hidden fields being in charge. So the change of ${\bf x}\to -{\bf x}$ in these four equalities of Eq.(8.43) does reverse the status of $\eta_c$ versus $\xi_c$ (or $\eta'_c$ vs $\xi'_c$), rendering helicity reversed explicitly. The subtlety of tachyon equation, unlike Dirac equation, lies in the fact that only $\nu_L$ and $\bar{\nu}_R$ exist whereas $\nu_R$ and $\bar{\nu}_L$ are strictly forbidden, \ie, the parity symmetry is violated to maximum. Hence, in strict sense, there is also no physically meaningful WF after the operation of pure "time inversion" on Eq.(8.15). We will insist on Eq.(8.31) rather than Eq.(8.42) --- there is only one correct way leading from tachyon to antitachyon via the ${\cal PT}$ inversion essentially.

In 2000, Eq.(8.25) was first proposed by Chang and then collaborated with Ni in Ref.\cite{39} (see also \cite{40,41,42,43} and the Appendix 9B in Ref.\cite{20}). At first sight, the difference between Eqs.(8.25) and (8.1) amounts to substituting the mass term $\beta m$ by $\beta_s m_s$ with $\beta_s=\begin{pmatrix}0& I\\ -I& 0\end{pmatrix}$ being an antihermitian matrix. Usually, for an equation with nonhermitian Hamiltonian, there is no guarantee for the completeness of its mathematical solutions. In other words, the unitarity of its physical states is at risk. Sometimes, however, a nonhermitian Hamiltonian can be accepted in physics. For example, in the optical model for nuclear physics, an imaginary part of potential, $V=V_0+iV_1$, is used to describe the absorption of incident particles successfully. The interesting thing for "tachyonic neutrino" is: Solutions of Eq.(8.15) (or (8.26)) for $E>0$ ($E_c>0$) are coinciding with that for $|\xi|>|\eta|$ ($|\eta_c|>|\xi_c|$) whereas another would-be solutions with $E>0$ but $|\xi|<|\eta|$ ($E_c>0$ but $|\eta_c|<|\xi_c|$) are forbidden, see Eqs.(8.37) and (8.39). It seems like half of would-be solutions disappear automatically. Equivalently, from physical point of view, only half of states with $\rho>0$ or $\rho_c>0$ are allowed in nature whereas another half with $\rho<0$ or $\rho_c<0$ are not. Hence one unique feature of "tachyon" equation, like Eq.(8.15) or (8.26), lies in its strange realization of unitarity violation that half of would-be states (being tentatively identified with $\nu_R$ and $\bar{\nu}_L$) are absolutely forbidden whereas another half ($\nu_L$ and $\bar{\nu}_R$) are stabilized. The permanently longitudinal polarization property of neutrino and antineutrino like that analysed above was first predicted by Lee and Yang in 1957 \cite{3} and had been verified by GGS experiment in 1958 \cite{44}. Further discussion on this topic is currently in preparation.

\section{IX. Antigravity Between Matter and Antimatter}
\label{sec:discussion}
\setcounter{equation}{0}
\renewcommand{\theequation}{9.\arabic{equation}}

In hindsight, there are two Lorentz invariants in the kinematics of SR:
\begin{equation}\label{6-1}
c^2(t_1-t_2)^2-({\bf x}_1-{\bf x}_2)^2=c^2(t'_1-t'_2)^2-({\bf x'}_1-{\bf x'}_2)^2=const
\end{equation}
\begin{equation}\label{6-2}
E^2-c^2{\bf p}^2=E'^2-c^2{\bf p'}^2=m^2c^4
\end{equation}
It seems quite clear that Eq.(\ref{6-1}) is invariant under the space-time inversion (${\bf x}\to-{\bf x},t\to-t$) and Eq.(\ref{6-2}) remains invariant under the mass inversion ($m\to-m$). We believe that these two discrete symmetries are deeply rooted at the SR's dynamics via its combination with QM and developing into RQM and QFT --- the particle and its antiparticle are treated on equal footing and linked by the symmetry ${\cal P}{\cal T}={\cal C}$ essentially. Hence we can perform a mass inversion on Eq.(9.2) in each of two inertial frames with arbitrary relative velocity $v$ in the sense of $m\to-m_c=-m, E\to-E_c, {\bf p}\to-{\bf p}_c$, yielding:
\begin{equation}\label{}
E_c^2-c^2{\bf p}_c^2={E'}_c^2-c^2{\bf p'}_c^2=m_c^2c^4=m^2c^4
\end{equation}
The invariance of Eq.(9.2) under mass inversion as a whole reflects the experimental fact that particle and antiparticle are equally existing in nature even at the level of classical physics.

Example: The motion equation for a charged particle (say, electron with charge $q=-e<0$) in the external electric and magnetic fields, ${\bf E}$ and ${\bf B}$, is given by the Lorentz formula:
\begin{equation}\label{}
m{\boldsymbol a}=q\left({\bf E}+\frac{1}{c}{\bf v}\times{\bf B}\right)
\end{equation}
Then the operation of either $q\to q_c=-q$ or $m\to-m_c=-m$ on Eq.(9.4) will realize the transformation from particle into its antiparticle (say, positron with charge $q_c=-q=e>0$) with the acceleration change from ${\bf a}\to {\bf a}_c=-{\bf a}$ as
\begin{equation}\label{}
m{\boldsymbol a}_c=-q\left({\bf E}+\frac{1}{c}{\bf v}\times{\bf B}\right)
\end{equation}
Based on what we learn from RQM (sections III-V) as well as Eqs.(9.1)-(9.5), we may conjecture that for a classical theory being capable of treating matter and antimatter on an equal footing, it must be invariant under a mass inversion $m\to-m_c=-m$.

Notice that, however, Eq.(9.4) (Eq.(9.5)) is only valid for particle (antiparticle) moving at low speed, it must be modified to adapt to high-speed cases through the invariance of continuous Lorentz transformation. So we need "double checks" for testing a classical theory being really "relativistic" or not.

Let us restudy the theory of general relativity (GR). In a $(-,+,+,+)$ metric, the Einstein field equation (EFE) reads (see, \eg, Refs.\cite{45,46}), ($c=1$),
\begin{equation}\label{}
G_{\mu\nu}\equiv R_{\mu\nu}-\frac{1}{2}g_{\mu\nu}R=-8\pi GT_{\mu\nu}
\end{equation}
Of course, Eq.(9.6) is covariant with respect to the Lorentz transformation. But could it withstand the test of mass inversion?

On the LHS of Eq.(9.6), the Einstein tensor $G_{\mu\nu}$ contains no any mass and no charge as well. But on the RHS, the energy-momentum current density tensor $T_{\mu\nu}$ is proportional to particle's mass $m$ and so changes its sign under an operation of $m\to-m$. Hence as a whole, Eq.(9.6) cannot remain invariant under the mass inversion. The reason seems rather clear that antimatter was not taking into account when GR was established in 1915. To modify EFE such that it can preserve the invariance of mass inversion, in 2004, one of us (Ni) proposed to add another term with $T_{\mu\nu}^c$ for antimatter, yielding
\begin{equation}\label{}
R_{\mu\nu}-\frac{1}{2}g_{\mu\nu}R=-8\pi G(T_{\mu\nu}-T_{\mu\nu}^c)
\end{equation}
which remains invariant under a mass inversion since:
\begin{equation}\label{}
T_{\mu\nu}\to-T_{\mu\nu}^c,\,T^c_{\mu\nu}\to-T_{\mu\nu}\qquad (\,m\to-m)
\end{equation}
In a weak-field (or the post-Newtonian) approximation, this modified EFE, MEFE, Eq.(9.7), will lead to modified Newton gravitational law as
\begin{equation}\label{}
F_{grav}(r)=\mp G\dfrac{mm'}{r^2}
\end{equation}
where the "$-$" sign means attractive force between $m$ and $m'$ being both matter or antimatter whereas the "$+$" sign means repulsive force between $m$ and $m'$ (both positive) if one of them is antimatter.

If we define the "gravitational mass" for matter and antimatter separately
\begin{equation}\label{}
m_{grav}=\left\{\begin{array}{ll}
          m>0,& (\text{matter}) \\
         -m_c=-m<0& (\text{antimatter})
         \end{array}\right.
\end{equation}
Then Eq.(9.9) can be recast into one equation
\begin{equation}\label{}
F_{grav}(r)=-G\dfrac{m_{grav}m'_{grav}}{r^2}
\end{equation}
which bears a close resemblance to the Coulomb law in classical electrodynamics (CED)
\begin{equation}\label{}
F_{Coul}(r)=\dfrac{qq'}{r^2}
\end{equation}
In 1986, within the framework of classical field theory (CFT) plus some assumptions, Jagannathan and Singh derived the potential energy of two static point sources as \cite{47}
\begin{equation}\label{}
U(r)=(-1)^{n+1}ee'\times(\text{a positive number})\times\dfrac{e^{-\mu r}}{r}
\end{equation}
where $n$ and $\mu$ are spin and mass of the mediating field, $e$ is the "charge" of the source. For CED, $n=1$ whereas $n=2$ for gravitational field ($\mu\to0$ in both cases). So Eq.(9.13) is in conformity with Eqs.(9.11) and (9.12) for the case of "like sources" (with $ee'>0$) \cite{47}, where the case for "unlike sources" ($ee'<0$) hadn't been discussed. Here Eq.(9.11) has been generalized to the case for "unlike sources", but at a price that the "equivalence principle" in GR ceases to be valid when matter and antimatter coexist as shown by Eq.(9.10).

In 2011, the antigravity between matter and antimatter was also claimed by Villata in Ref.\cite{48}, where the argument seems different from that explained above. But theory is theory, only fact will have the final say. So we are anxiously waiting for the outcome from the AEGIS experiment \cite{49} (at CERN), which is designed to compare the Earth gravitational acceleration on hydrogen and antihydrogen atoms.

\section{X. Summary}
\label{sec:discussion}
\setcounter{equation}{0}
\renewcommand{\theequation}{10.\arabic{equation}}

1. Being the combination of SR and QM, RQM is capable of dealing with particle and antiparticle on an equal footing. As long as we admit that the antiparticle's momentum and energy operators should be $\hat{\bf p}_c=i\hbar\nabla$ and $\hat{E}_c=-i\hbar\frac{\partial}{\partial t}$ versus $\hat{\bf p}=-i\hbar\nabla$ and $\hat{E}=i\hbar\frac{\partial}{\partial t}$ for particle, it can be proved that the "negative-energy" WF $\psi$ of particle corresponds to a "positive-energy" WF $\psi_c$ of antiparticle precisely.

2. In general, an equation in RQM always has a discrete symmetry ${\cal P}{\cal T}={\cal C}$ which shows up as a transformation between a particle's WF $\psi$ and its antiparticle's WF $\psi_c$: $\psi({\bf x},t)\rightleftarrows\psi_c({\bf x},t)$. For a free particle, it simply means $\psi(-{\bf x},-t)=\psi_c({\bf x},t)$. This is in conformity with the "strong reflection" in QFT invented by Pauli and L\"{u}ders, showing that the intrinsic property of a particle cannot be detached from the space-time.

3. Following Feshbach-Villars' deep insight, we are able to divide each and every WF $\psi$ in RQM into two parts, $\psi=\phi+\chi$. Then the above symmetry is further rigorously expressed by an invariance of motion equation in RQM through the transformations $\phi\rightleftarrows\chi_c$ and $\chi\rightleftarrows\phi_c$ under either the space-time inversion (${\bf x}\to -{\bf x}, t\to-t$) or a mass inversion ($m\to -m$). Since $|\phi|>|\chi|$ in $\psi$ whereas $|\chi_c|>|\phi_c|$ in $\psi_c$, we may name $\phi$ as the (dominant) hidden particle field in $\psi$ while $\chi$ the (subordinate) hidden antiparticle field in $\psi$. In this way, both the "probability density" $\rho$ for a particle and $\rho_c$ for an antiparticle can be proved to be positive definite. Now we may say that the RQM is ensured to be self-consistent and can be regarded as a sound basis for QFT.

4. All kinematical effects in SR can be ascribed to the enhancement of the magnitude of $\chi$ field in a particle's WF accompanying with the increase of particle's velocity.

5. As proved for Dirac particle with spin, the helicity of a particle is just opposite to that of its antiparticle under a space-time (or mass) inversion. Therefore, the experimental tests for the CPT invariance should include not only the equal mass and lifetime of particle versus antiparticle, but also the following fact: A particle and its antiparticle with opposite helicities must coexist in nature with no exception. A prominent example is the neutrino --- A neutrino $\nu_L$ (antineutrino $\bar{\nu}_R$) is permanently left-handed (right-handed) polarized whereas the fact that no $\nu_R$ exists in nature must means no $\bar{\nu}_L$ as well (as verified by the GGS experiment \cite{44}). See also section VII.

6. Based on the invariance of space-time inversion or mass inversion (at the level of RQM) and the latter's generalization to the classical physics, we tentatively discuss some interesting problems in today's physics, including the prediction of antigravity between matter and antimatter, as well as the reason why we believe neutrinos are likely the tachyons.

\section*{Appendix: Klein Paradox for Klein-Gordon Equation and Dirac Equation}

We will discuss the Klein paradox \cite{50} for both KG equation and Dirac equation based on sections III and V, without resorting to the "hole" theory.

\subsection*{AI: Klein Paradox for KG Equation}

Consider that a KG particle moves along $z$ axis in one-dimensional space and hits a step potential
\begin{equation*}
V(z)=\left\{
       \begin{array}{ll}
         0, & \hbox{$z<0$;} \\
         V_0, & \hbox{$z>0$.}
       \end{array}
     \right.\eqno{(A.1)}
\end{equation*}
Its incident WF with momentum $p\,(>0)$ and energy $E\,(>0)$ reads
\begin{equation*}
\psi_i=a\exp[i(pz-Et)],\quad (z<0)\eqno{(A.2)}
\end{equation*}
If $E=\sqrt{p^2+m^2}<V_0$, we expect that the particle wave will be partly reflected at $z=0$ with WF $\psi_r$ and another transmitted wave $\psi_t$ emerged at $z>0$:
\begin{equation*}
\psi_r=b\exp[i(-pz-Et)],\quad (z<0)\eqno{(A.3)}
\end{equation*}
\begin{equation*}
\psi_t=b'\exp[i(p'z-Et)],\quad (z>0)\eqno{(A.4)}
\end{equation*}
with $p'^2=(E-V_0)^2-m^2$. See Fig.1(a).
\begin{figure}
  \hspace*{-3mm}\includegraphics[scale=0.98]{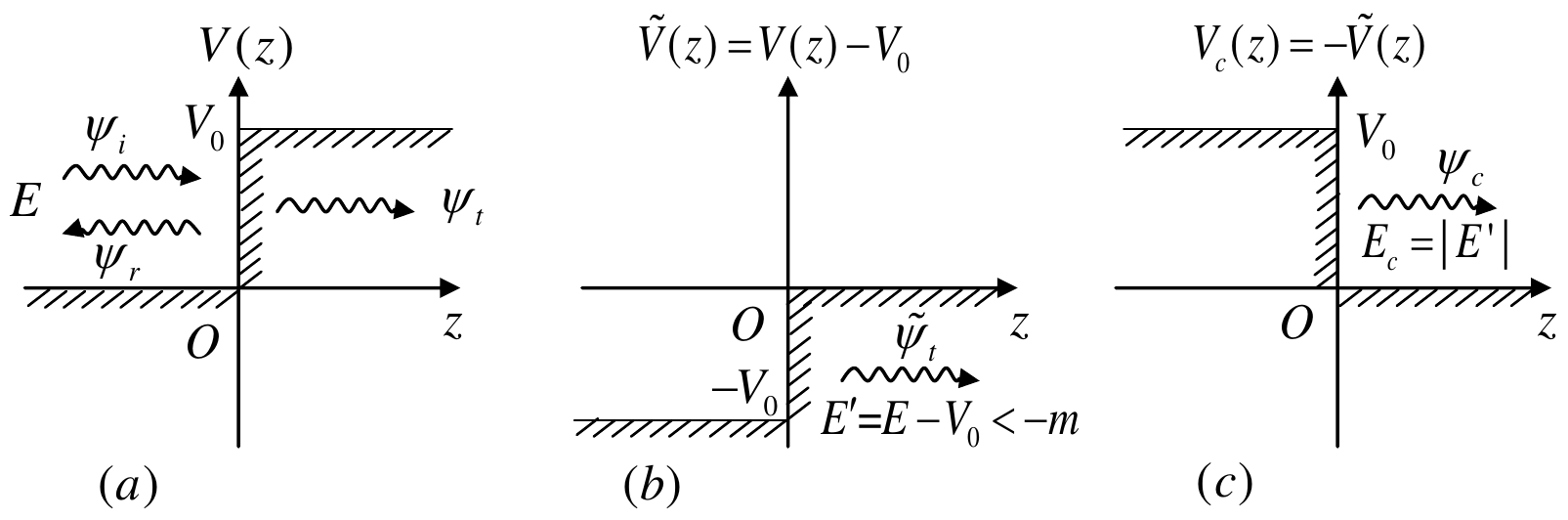}\\
  \caption{Klein paradox: (a) If $V_0>E+m$, there will be a wave $\psi_t$ at $z>0$.\\ (b) Just look at $z>0$ region, making a shift $V(z)\to\tilde{V}(z)=V(z)-V_0,\,E\to E'=E-V_0<-m$.\\ (c) An antiparticle (at $z>0$) appears with its energy $E_c=|E'|>m$ and the potential is $V_c(z)=-\tilde{V}(z)$}
\end{figure}

Two continuity conditions for WFs and their space derivatives at the boundary $z=0$ give two simple equations
\begin{equation*}
\left\{\begin{array}{l}
         a+b=b' \\
        (a-b)p=b'p'
       \end{array}\right.
\eqno{(A.5)}
\end{equation*}
The Klein paradox happens when $V_0>E+m$ because the momentum $p'=\pm\sqrt{(V_0-E)^2-m^2}$ is real again and the reflectivity $R$ of incident wave reads
\begin{equation*}
R=\left|\dfrac{b}{a}\right|^2=\left|\dfrac{p-p'}{p+p'}\right|^2,\;\left\{\begin{array}{l}
         R<1,\quad\text{if}\;p'>0 \\
        R>1,\quad\text{if}\;p'<0
       \end{array}\right.
\eqno{(A.6)}
\end{equation*}
(See Ref.\cite{14} or \S9.4 in Ref.\cite{20}, where discussions are not complete and need to be complemented and corrected here). Because the kinetic energy $E'$ at $z>0$ is negative: $E'=E-V_0<0$, what does it mean? Does the particle still remain as a particle?

As discussed in section III, for a KG particle (or its antiparticle), two criterions must be held: its probability density $\rho$ (or $\rho_c$) must be positive and its probability current density ${\bf j}$ (or ${\bf j}_c$) must be in the same direction of its momentum ${\bf p}$ (or ${\bf p}_c$).

See Fig.1(b), after making a shift in the energy scale, \ie, basing on the new vacuum at $z>0$ region, we redefine a WF $\tilde{\psi}_t$ (which is actually the WF in the "interaction picture", $\tilde{\psi}_t=\psi_te^{iV_0t}\,(z>0)$)
\begin{equation*}
\psi_t\to\tilde{\psi}_t=b'\exp[i(p'z-E't)],\quad (z>0)\eqno{(A.7)}
\end{equation*}
($E'=E-V_0<0$). From now on we will replace KG WF $\tilde{\psi}_t$ by $\tilde{\phi}_t$ and $\tilde{\chi}_t$ according to Eq.(\ref{55}), if $\tilde{\psi}_t$ still describes a "particle", whose probability density $\rho_t$ should be evaluated by Eq.(\ref{56}) with $V\to\tilde{V}(z)=0\,(z>0)$ yielding:
\begin{equation*}
\rho_t=|\tilde{\phi}_t|^2-|\tilde{\chi}_t|^2=\dfrac{E'}{m}|b'|^2<0,\quad (z>0)\eqno{(A.8)}
\end{equation*}
And its probability current density $j_t$ should be given by Eq.(\ref{53}), yielding:
\begin{equation*}
j_t=\dfrac{p'}{m}|b'|^2,\quad (z>0)\eqno{(A.9)}
\end{equation*}
Eq.(A.8) is certainly not allowed. So to consider a "particle" with momentum $p'>0$ moving to the right makes no sense. Instead, we should consider $p'<0$ (which also makes no sense for a particle due to the boundary condition) and regard $\tilde{\psi}_t$ as an antiparticle's WF by rewriting it as:
\begin{equation*}
\tilde{\psi}_t=\psi_c=b'\exp[-i(p_cz-E_ct)],\quad (z>0)\eqno{(A.10)}
\end{equation*}
Now using Eq.(2.18) we see that Eq.(A.10) does describe an antiparticle with momentum $p_c=-p'=|p'|=\sqrt{E_c^2-m^2}>0$ and energy $E_c=|E'|=V_0-E>0$. In the mean time, from the antiparticle's point of view (\ie, with $E_c>m$), the potential becomes $V_c(z)=-\tilde{V}(z)$ (comparing Eq.(2.21) with Eq.(A.10) as shown by Fig.1(c).

It is easy to see from Eqs.(3.30),(3.31) and (A.10) that\footnotemark[1]\footnotetext[1]{We had discarded the solution of $p'>0$ in Eqs.(A.7)-(A.9) as a particle. However, if we consider $p'=-p_c>0$ for an antiparticle, then similar to Eqs.(A.10)-(A.11), we would get $\rho_t^c>0$ but both $j_t^c$ and $p_c$ are negative, meaning that the antiparticle is coming from $z=\infty$, not in accordance with our boundary condition. So the case of $p'>0$ should be abandoned either as a particle or as an antiparticle.}
\begin{equation*}
\left\{\begin{array}{l}
       \rho_t^c=|\tilde{\chi}_t^c|^2-|\tilde{\phi}_t^c|^2=\dfrac{E_c}{m}|b'|^2>0,   \\[4mm]
       j_t^c=\dfrac{p_c}{m}|b'|^2
       \end{array}\right.
\quad (z>0)\eqno{(A.11)}
\end{equation*}
So the reflectivity, Eq.(A.6), should be fixed as:
\begin{equation*}
R_{KG}=\left|\dfrac{b}{a}\right|^2=\left|\dfrac{p+p_c}{p-p_c}\right|^2=\left(\dfrac{1+\gamma'}{1-\gamma'}\right)^2,\;
\gamma'=\dfrac{p_c}{p}>0 \eqno{(A.12)}
\end{equation*}
And the transmission coefficient can also be predicted as:
\begin{equation*}
T_{KG}=\dfrac{j_t^c}{j_i}=\dfrac{p_c}{p}\left|\dfrac{b'}{a}\right|^2=\dfrac{p_c}{p}\left|1+\dfrac{b}{a}\right|^2
=\dfrac{4pp_c}{(p-p_c)^2}=\dfrac{4\gamma'}{(1-\gamma')^2}\eqno{(A.13)}
\end{equation*}
\begin{equation*}
 R_{KG}-T_{KG}=1\eqno{(A.14)}
\end{equation*}
The variation of $T_{KG}$ seems very interesting:
\begin{equation*}
T_{KG}=\left\{\begin{array}{l}
0,\;\gamma'\to 0\quad(p_c\to 0,E_c\to m) \\[3mm]
\infty,\;\gamma'\to 1\quad(p_c=p,E_c=E=V_0/2)\\[3mm]
0,\;\gamma'\to \infty\quad(p_c\to \infty,E_c=V_0-E\to \infty)\\[3mm]
0,\;\gamma'\to \infty\quad(p\to 0,E\to m)
              \end{array}\right.
\eqno{(A.15)}
\end{equation*}
Above equations show us that the incident KG particle triggers a process of "pair creation" occurring at $z=0$, creating new particles moving to the left side (to join the reflected incident particle) so enhancing the reflectivity $R_{KG}>1$ and new antiparticles (with equal number of new particles) moving to the right.

To our understanding, this is not a stationary state problem for a single particle, but a nonstationary creation process of many particle-antiparticle system. It is amazing to see the Klein paradox in KG equation being capable of giving some prediction for such kind of process at the level of RQM. Further investigations are needed both theoretically and experimentally. \footnotemark[2]\footnotetext[2]{We find from the Google search that R. G. Winter in 1958 had written a paper titled "Klein paradox for the Klein-Gordon equation" and reached basically the same result as ours. So he was the first author dealing with this problem. Regrettably, it seems that his paper had never been published on some journal.}

\subsection*{AII: Klein Paradox for Dirac Equation}

Beginning from Klein \cite{50}, many authors \eg~ Greiner \etal \cite{51,52}, have studied this topic. We will join them by using the similar approach like that for KG equation discussed above.

Based on similar picture shown in Fig.1, now we have three Dirac WFs under the condition $V_0>E+m$:
\begin{equation*}
\psi_i=a\begin{pmatrix}1\\0\\\frac{p}{E+m}\\0\end{pmatrix}
e^{i(pz-Et)},\psi_r=b\begin{pmatrix}1\\0\\\frac{-p}{E+m}\\0\end{pmatrix}
e^{i(-pz-Et)}\quad (z<0)\eqno{(A.16)}
\end{equation*}
\begin{equation*}
\psi_t=b'\begin{pmatrix}1\\0\\\frac{p'}{E-V_0+m}\\0\end{pmatrix}e^{i(p'z-Et)}
=b'\begin{pmatrix}1\\0\\\frac{-p'}{V_0-E-m}\\0\end{pmatrix}e^{i(p'z-Et)}
=\begin{pmatrix}\phi_t\\\chi_t\end{pmatrix}\quad (z>0)\eqno{(A.17)}
\end{equation*}
where $p'=\pm\sqrt{(V_0-E)^2-m^2}$. Unlike Eq.(A.8) for KG equation, the probability density for Dirac WF $\psi_t$ is positive definite (see Eq.(\ref{81}))
\begin{equation*}
\rho_t=\psi^\dag_t\psi_t=\phi^\dag_t\phi_t+\chi^\dag_t\chi_t\eqno{(A.18)}
\end{equation*}
Hence we will rely on two criterions: First, the probability current density and momentum must be in the same direction for either a particle or antiparticle. For $\psi_i$ and $\psi_r$, their probability current density are ($c=1$)
\begin{equation*}\begin{array}{l}
j_i=\psi^\dag_i\alpha_z\psi_i=\phi_i^\dag\sigma_z\chi_i+\chi^\dag_i\sigma_z\phi_i=\dfrac{2p}{E+m}|a|^2>0 \\
j_r=\psi^\dag_r\alpha_z\psi_r=\dfrac{-2p}{E+m}|b|^2<0
                 \end{array}\quad (z<0)
\eqno{(A.19)}
\end{equation*}
as expected. However, for $\psi_t$, we meet difficulty similar to that in Eq.(A.9)
\begin{equation*}
j_t=\psi^\dag_t\alpha_z\psi_t=\dfrac{-2p'}{V_0-E-m}|b'|^2 \quad (z>0)\eqno{(A.20)}
\end{equation*}
the direction of $j_t$ is always opposite to that of $p'$! The second criterion is: while $|\phi|>|\chi|$ for particle, we must have $|\chi_c|>|\phi_c|$ for antiparticle. Now in $\psi_i$ (or $\psi_r$), $|\phi_i|>|\chi_i|$ (or $|\phi_r|>|\chi_r|$), but the situation in $\psi_t$ is dramatically changed, the existence of $V_0$ renders $|\chi_t|>|\phi_t|$!

The above two criterions, together with the experience in KG equation, prompt us to choose $p'<0$ and regard $\psi_t$ as an antiparticle's WF. So we rewrite:
\begin{equation*}
\psi_t=\psi_t^ce^{-iV_0t}\eqno{(A.21a)}
\end{equation*}
\begin{equation*}
\psi_t^c=b'\begin{pmatrix}1\\0\\\frac{p_c}{E_c-m}\\0\end{pmatrix}
e^{-i(p_cz-E_ct)}=\begin{pmatrix}\phi_t^c\\\chi_t^c\end{pmatrix},
\tilde{\psi}^c_t=b'_c\begin{pmatrix}1\\0\\\frac{p_c}{E_c+m}\\0\end{pmatrix}
e^{-i(p_cz-E_ct)}=\begin{pmatrix}\chi_t^c\\\phi_t^c\end{pmatrix}
\quad (z<0)\eqno{(A.21b)}
\end{equation*}
where $\tilde{\psi}^c_t=(-\gamma^5)\psi_t^c$ (with new normalization constant $b'_c$ replacing
$b'$) describes an antiparticle with momentum $p_c=|p'|=-p'=\sqrt{E_c^2-m^2}>0$, energy $E_c=V_0-E>0$ and $|\chi_t^c|>|\phi_t^c|$. Using Eq.(\ref{82}) we find
\begin{equation*}
j_t^c=\dfrac{2p_c}{E_c+m}|b'_c|^2>0,\quad (z>0)\eqno{(A.22)}
\end{equation*}
as expected. Now it is easy to match Dirac WFs at the boundary $z=0$, ($\psi_i+\psi_r)|_{z=0}=\tilde{\psi}_t^c|_{z=0}$, yielding\footnotemark[1]\footnotetext[1]{Eq.(A.23) means that the large (small) component of spinor is connected with
large (small) component at both sides of $z=0$. However, if instead of $\tilde{\psi}^c_t$, the $\psi_t^c$ is used directly with its first (small) component being connected with the first (large) components of $\psi_i$ and $\psi_r$, it would lead to a different expression of Eq.(A.27): $\gamma\to\tilde{\gamma}=\sqrt{\frac{(E_c-m)(E-m)}{(E+m)(E_c+m)}}$, which is just the $1/\gamma$ ($\gamma$ and $1/\gamma$ make no difference in the result of, say, Eqs.(A.24) and (A.25)) defined by Eq.(8) on page 266 of Ref.\cite{51} (see Eq.(A31) below) or that by Eq.(5.36) in Ref.\cite{52}}
\begin{equation*}\left\{\begin{array}{l}
a+b=b'_c \\
\dfrac{(a-b)p}{E+m}=\dfrac{b'_cp_c}{E_c+m}
                 \end{array}\right.\to\left\{\begin{array}{l}
\dfrac{b}{a}=\dfrac{\xi-\eta}{\xi+\eta} \\
\dfrac{b'_c}{a}=1+\dfrac{b}{a}=\dfrac{2\xi}{\xi+\eta}
                 \end{array}\right.
\eqno{(A.23)}
\end{equation*}
where $\xi=p(E_c+m)>0,\eta=p_c(E+m)>0$. The reflectivity $R_D$ and transmission coefficient $T_D$ follow from Eq.(A.19) and (A.22) as:
\begin{equation*}
R_D=\dfrac{|j_r|}{j_i}=\left|\dfrac{b}{a}\right|^2=\left(\dfrac{1-\gamma}{1+\gamma}\right)^2\eqno{(A.24)}
\end{equation*}
\begin{equation*}
T_D=\dfrac{j_t^c}{j_i}=\left|\dfrac{b'_c}{a}\right|^2\dfrac{p_c(E+m)}{p(E_c+m)}=\dfrac{4\gamma}{(1+\gamma)^2}\eqno{(A.25)}
\end{equation*}
\begin{equation*}
 R_D+T_D=1\eqno{(A.26)}
\end{equation*}
where
\begin{equation*}
\gamma=\dfrac{\eta}{\xi}=\sqrt{\dfrac{(E_c-m)(E+m)}{(E-m)(E_c+m)}}\geq0\;(E_c=V_0-E\geq m)\eqno{(A.27)}
\end{equation*}
and
\begin{equation*}
T_D=\left\{\begin{array}{l}
0,\;\gamma\to 0\quad(p_c\to 0,E_c\to m) \\
1,\;\gamma= 1\quad(p_c=p,E_c=E=V_0/2)\;(\text{resonant\ transmission}) \\
\frac{2p}{E+p},\;\gamma\to \sqrt{\frac{E+m}{E-m}}\quad(E_c=V_0-E\to \infty)\\
0,\;\gamma\to \infty\quad(p\to 0,E\to m)
              \end{array}\right.
\eqno{(A.28)}
\end{equation*}
The variation of $T_D$ bears some resemblance to Eq.(A.15) for KG equation but shows striking difference due to sharp contrast between Eqs.(A.24)-(A.28) and Eqs.(A.12)-(A.15).

To our understanding, in the above Klein paradox for Dirac equation, there is no "pair creation" process occurring at the boundary $z=0$. The paradox just amounts to a steady transmission of particle's wave $\psi_i$ into a high potential barrier $V_0>E+m$ at $z>0$ region where $\psi_t$ shows up as an antiparticle's WF propagating to the right. In some sense, the existence of a potential barrier $V_0$ plays a "magic" role of transforming the
particle into its antiparticle. Because the probability densities of both particle and antiparticle are positive definite, the total probability can be normalized over the entire space like that for one particle case:
\begin{equation*}
\int_{-\infty}^\infty[\rho(z)\Theta(-z)+\rho_c(z)\Theta(z)]dz=1\eqno{(A.29)}
\end{equation*}
($\Theta(z)$ is the Heaviside function) and the probability current density remains continuous at the boundary $z=0$. In other words, the continuity equation holds in the whole space just like what happens in a one-particle stationary state.

It is interesting to compare our result with that in Refs.\cite{51} and \cite{52}. In Ref.\cite{51}, Eqs.(13.24)-(13.28) are essentially the same as ours. But the argument there for choosing $\bar{p}<0$ in Eq.(13.23) is based on the criterion of the group velocity $v_{gr}$ being positive (for the transmitted wave packet moving toward $z=\infty$). And the $v_{gr}$ is stemming from Eq.(13.16) which is essentially the probability current density in our Eqs.(A.21)-(A.22).

However, the author in Ref.\cite{51} also considered the other choice $\bar{p}>0$ in an example (p.265-267 in \cite{51}) based on the hole theory, ending up with the prediction as:
\begin{equation*}
R=\left(\dfrac{1+\gamma}{1-\gamma}\right)^2,\;T=\dfrac{4\gamma}{(1-\gamma)^2},\;R-T=1\eqno{(A.30)}
\end{equation*}
where
\begin{equation*}
\gamma=\dfrac{p_2}{p_1}\dfrac{E+m}{V_0-E-m}=\sqrt{\dfrac{(V_0-E+m)(E+m)}{(V_0-E-m)(E-m)}}\eqno{(A.31)}
\end{equation*}
The argument for the validity of his Eqs.(A.30)-(A.31) is based on the hole theory (see also section 5.2 in Ref.\cite{52}), saying that once $V_0>E+m$, there would be an overlap between the occupied negative continuum for $z>0$ and the empty positive continuum for $z<0$, providing a mechanism for electron-positron pair creation if the "hole" at $z>0$ can be identified with a positron. We doubt the "hole" theory seriously because there are only two electrons (with opposite spin orientations) staying at each energy level in the negative continuum. So it seems that there is no abundant source for electrons and "holes" to account for the huge value of $T>1$ in Eq.(A.30).

Fortunately, we learn from section 10.7 in Ref.\cite{52} that if the Klein paradox in Dirac equation is treated at the level of QFT, their result turns out to be the same form as our Eqs.(A.24)-(A.28), rather than Eqs.(A.30)-(A.31).

\vspace*{-2mm}
\section*{Acknowledgements}
\vspace*{-2mm}
We thank E. Bodegom, T. Chang, Y. X. Chen, T. P. Cheng, X. X. Dai, G. Tananbaum, V. Dvoeglazov, Y. Q. Gu, F. Han, J. Jiao, A. Kellerbauer, T. C. Kerrigan, A. Khalil, R. Konenkamp, D. X. Kong, J. S. Leung, P. T. Leung, Q. G. Lin, S. Y. Lou, D. Lu, Z. Q. Ma, D. Mitchell, E. J. Sanchez, Z. Y. Shen, Z. Q. Shi, P. Smejtek, X. T. Song, R. K. Su, Y. S. Wang, Z. M. Xu, X. Xue, J. Yan, F. J. Yang, J. F. Yang, R. H. Yu, Y. D. Zhang and W. M. Zhou for encouragement, collaborations and helpful discussions.




\begin{thebibliography}{99}

\bibitem[1]{1}
A. Apostolakis \etal (CPLEAR Collaboration), "An EPR experiment testing the non-separability of the $K^0\bar{K}^0$ wave function", Physics Letters B, Vol.422, 1998, pp339-348.

\bibitem[2]{2}
H. Feshbach and F. Villars, "Elementary relativistic wave mechanics of spin-0 and spin-1/2 particles", Review of Modern Physics, Vol.30, 1958, pp24-45.

\bibitem[3]{3}
T. D. Lee and C. N. Yang, "Question of Parity Conservation in Weak Interactions", Physical Review, Vol.104, No.1, 1956, pp254-258;  "Parity Nonconservation and a Two-Component Theory of the Neutrino", {\it ibid}, Vol.105, No.5, 1957, pp1671-1675; T. D. Lee, R. Oehme and C. N. Yang, "Remarks on Possible Noninvariane under Time Reversal and Charge Conjugation", {\it ibid}, Vol.106, No.2, 1957, pp340-345.

\bibitem[4]{4}
C. S. Wu, E. Ambler, R. W. Hayward, D. D. Hoppes and R. P. Hudson, "Experimental Test of Parity Conservation in Beta Decay", Physical Review, Vol.105, No.4, 1957, pp1413-1415.

\bibitem[5]{5}
J. H. Christensen, J. W. Cronin, V. L. Fitch and R. Turlay, "Evidence for the 2~$\pi$ Decay of the $K_2^0$ Meson", Physical Review Letters, Vol.13, No.4, 1964, pp138-140.

\bibitem[6]{6}
K. R. Schubert, B. Wolff, J.-M. Gaillard, M.R. Jane, T.J. Ratcliffe, J.-P. Repellin, "The phase of $\eta_{00}$ and the invariances CPT and T", Physics Letters B, Vol.31, No.10, 1970, pp662-665.

\bibitem[7]{7}
J. Beringer \etal~(Particle Data Group), "Review of Particle Physics", Physical Review D, Vol.86, No.1, 2012, 010001 [1528 pages].

\bibitem[8]{8}
G. L\"{u}ders, "On the equivalence of invariance under time reversal and under particle-antiparticle conjugation for relativistic field theories", Kgl. Danske Vidensk. Selsk. Mat.-Fys. Medd. Vol.28, No.5, 1954; "Proof of the TCP Theorem", Annals of Physics (New York), Vol.2, 1957, pp1-15.

\bibitem[9]{9}
W. Pauli, "Exclusion principle, Lorentz group and reflection of space-time and charge" in {\it Niels Bohr and the Development of Physics}, ed. by W. Pauli, L. Rosenfeld and V. Weisskopf (McGraw-Hill, 1955), pp30-51.

\bibitem[10]{10}
T. D. Lee and C. S. Wu, "Weak Interactions", Annual Review of Nuclear Science, Vol.15, 1965, pp381-476.

\bibitem[11]{11}
A. Einstein, B. Podolsky and N. Rosen, "Can Quantum -Mechanical Description of Physical Reality Be Considered Complete ?", Physical Review, Vol.47, No.10, 1935, pp777-780.

\bibitem[12]{12}
D. Bohm, {\it Quantum Theory}, Prentice Hall, 1956; J. S. Bell, "On the Einstein Podolsky Rosen Paradox", {\it Physics}, Long Island City, NewYork, Vol.1, No.3, 1964, pp195-200.

\bibitem[13]{13}
H. Guan, {\it Basic Concepts in Quantum Mechanics}, (High Education Press, Beijing, 1990), Chapter 7 (in Chinese).

\bibitem[14]{14}
G. J. Ni, H. Guan, W. M. Zhou and J. Yan, "Antiparticle in Light of Einstein-Podolsky-Rosen Paradox and Klein Paradox", Chinese Physics Letters, Vol.17, 2000, pp393-395, {\tt quant-ph}/0001016.

\bibitem[15]{15}
O. Nachtmann, {\it Elementary Particle Physics, Concepts and Phenomena}, Springer-Verlag, 1990.

\bibitem[16]{16}
W. Greiner and B. M\"{u}ller, {\it Gauge Theory of Weak Interactions}, Springer-Verlag, 1993, Ch.8.

\bibitem[17]{17}
E. J. Konopinski and H. M. Mahmaud, "The Universal Fermi Interaction", Physical Review, Vol.92, No.4, 1953, pp1045-1049 .

\bibitem[18]{18}
G. J. Ni, "Relation between space-time inversion and particle-antiparticle symmetry", Journal of Fudan University (Natural Science), No.3-4, 1974, pp125-134 (In fact, this paper was in collaboration with Suqing Chen, but her name was erased according to Editor's advice for promoting the publication at that time).

\bibitem[19]{19}
G. J. Ni and S. Q. Chen, "On the essence of special relativity", Journal of Fudan University (Natural Science)
Vol.35, No.3, 1996, pp325-334; English version titled "Relation between space-time inversion and particle-antiparticle symmetry and the Microscopic essence of special relativity", Ed. V. Dvoeglazov (NOVA Science Publisher, Inc. 1999), Chapter III, pp145-169; {\tt hep-th}/9508069.

\bibitem[20]{20}
G. J. Ni and S. Q. Chen, {\it Advanced Quantum Mechanics}, 2nd Edition (Fudan University Press, 2003); English Edition was
published by Rinton Press, 2002.

\bibitem[21]{21}
G. J. Ni, "Ten arguments for the essence of special relativity", Progress in Physics (Nanjing, China), Vol.23, No.4, 2003, pp484-503, (In English).

\bibitem[22]{22}
G. J. Ni, "A new insight into the negative-mass paradox of gravity and the accelerating universe" in {\it Relativity, Gravitation, Cosmology}, Edit by V. V. Dvoeglazov and A. A. Espinoza Garrido,
NOVA Science Publisher, 2004, pp123-136, \physics{0308038}.

\bibitem[23]{23}
G. J. Ni, J. J. Xu and S. Y. Lou, "Reduced Dirac equation and Lamb shift as off-mass-shell effect in quantum electrodynamics", Chinese Physics B, Vol.20, 2011, 020302, pp1-23; {\tt quant-ph}/0511197.

\bibitem[24]{24}
J. J. Sakurai, {\it Advanced Quantum Mechanics}, Addison-Wesley Publishing Company, 1978.

\bibitem[25]{25}
J. J. Sakurai, {\it Modern Quantum Mechanics}, NewYork, John Wiley \& Sons, Inc. 1994.

\bibitem[26]{26}
J. D. Bjorken and S. D. Drell, {\it Relativistic Quantum Mechanics}, McGraw-Hill. 1964, {\it Relativistic Quantum Field}, McGraw-Hill. 1967.

\bibitem[27]{27}
L. B. Okun, "The concept of mass", Physics Today, Vol.42, June 1989, pp31-36; Discussions in
Vol.42, May 1990, pp13, 15, 115, 117.

\bibitem[28]{28}
G. Lochak, "De Broglie's initial conception of De Broglie waves", in {\it The Wave-particle Dualism} (S. Diner \etal~ Eds., D. Reidel Publishing Company, 1984), pp1-25.

\bibitem[29]{29}
G. J. Ni, W. M. Zhou and J. Yan, "Comparison among Klein-Gordon equation, Dirac equation and relativistic Schr\"{o}dinger equation" in {\it Lorentz Group, CPT and Neutrinos}, Eds.: A. E. Chubykalo, V. V. Dvoeglazov, D. J. Ernst, V. G. Kadyshevsky, Y. S. Kim (World Scientific, 2000), pp.68-81.

\bibitem[30]{30}
M. E. Peskin and D. V. Schroeder, {\it An Introdution to Quantum Field Theory}, Addison-Wesley Publishing Company, 1995.

\bibitem[31]{31}
M. Jacob and G. C. Wicks, "On the General Theory of Collisions for Particles with Spin", Annals of Physics (New York), Vol.7, No.4, 1959, pp404-428 and references therein.

\bibitem[32]{32}
S. Weinberg, "A Model of Leptons", Physical Review Letters, Vol.19, No.21, 1967, pp1264-1266.

\bibitem[33]{33}
Z. Q. Shi and G. J. Ni, "Lifetime of polarized fermions in flight", Chinese Physics Letters, Vol.19, No.10, 2002, pp1427-1429.

\bibitem[34]{34}
Z. Q. Shi and G. J. Ni, "Calculations on the lifetime of polarized muons in flight", Annales de la Fondation Louis de Bloglie, Vol.29, Hors serie 2, 2004, pp1057-1066.


\bibitem[35]{35}
Z. Q. Shi and G. J. Ni, "The lifetime asymmetry of polarized fermions in flight", Handronic Journal, Vol.29, 2006, pp401-407; in {\it Frontiers in Horizons in World Physics}, Ed. Victor H. Marselle (Nova Science, 2008), pp.53-65.

\bibitem[36]{36}
Z. Q. Shi and G. J. Ni, "Experimental tests on the lifetime asymmetry", Modern Physics Letters A, vol.26, No13, 2011, pp987-998.

\bibitem[37]{37}
A. Cho, "Seeking a Shortcut to the High-Energy Frontier", Science, Vol.326, Dec 4, 2009, pp1342-1343.

\bibitem[38]{38}
L. H. Ryder, {\it Quantum Field Theory}, Cambridge University Press, Cambridge, 1996.

\bibitem[39]{39}
T. Chang and G. J. Ni, "An explanation of possible negative mass-square of neutrinos", FIZIKA (Zagreb), Vol.11, No.1, 2002, pp49-56, {\tt hep-ph}/0009291.

\bibitem[40]{40}
G. J. Ni and T. Chang, "Two parameters describing a superluminal neutrino", Journal of Shaanxi Normal University ( Natural Science Edition) Vol.30, No.3, 2002, pp32-39, {\tt hep-ph}/0103051.

\bibitem[41]{41}
G. J. Ni, "There might be superluminal particles in nature", {\it ibid}, Vol.29, No.3, 2001, pp1-5, {\tt hep-th}/0201077; "Superluminal paradox and neutrino", {\it ibid}, Vol.30, No.4, 2002, pp1-6, {\tt hep-th}/0203060.

\bibitem[42]{42}
G. J. Ni, "A minimal three flavor model for neutrino oscillation based on superluminal property", see Ref.[22], 2004, pp137-148, {\tt hep-ph/0306028}

\bibitem[43]{43}
G. J. Ni, "Principle of Relativity in Physics and in Epistemology" in {\it Relativity, Gravitation, Cosmology: New Development}, Editor: V. Dvoeglazov, (NOVA Science Publisher, 2010), pp237-252, {\tt physics/0407092}; "Cosmic Ray Spectrum and Tachyonic Neutrino", {\it ibid}, pp253-265, {\tt hep-ph/0404030}.

\bibitem[44]{44}
M. Goldhaber, L. Grodgins and A. W. Sunyar, "Helicity of Neutrinos", Physical Review, Vol.109, No.3, 1958, pp1015-1017.

\bibitem[45]{45}
S. Weinberg, {\it Gravitation and Cosmology}, John Wiley, 1972.

\bibitem[46]{46}
Z. M. Xu and X. J. Wu, {\it General Relativity and Contemporary Cosmology}, Press of Nanjing Normal University, 1999 (in Chinese); T. P. Cheng, {\it Relativity, Gravitation and Cosmology}, 2nd Edition, Oxford University Press, 2010.

\bibitem[47]{47}
K. Jagannathan and L. P. S. Singh, "Attraction/repulsion between like charges and the spin of the classical mediating field", Physical Review D, Vol.33, No.8, 1986, pp2475-2477.

\bibitem[48]{48}
M. Villata, "CPT symmetry and antimatter gravity in general relativity", Europhysics Letters, Vol.94, March 28, 2011, 20001 (pp1-6).

\bibitem[49]{49}
A. Kellerbauer \etal, "Proposed antimatter gravity measurement with an antihydrogen beam", Nuclear Instruments and Methods in Physics Research Section B, Vol.266, 2008, pp351-356.

\bibitem[50]{50}
O. Klein, Die Reflexion von Eleckronen an einem Potentialsprung nach der relativisticchen Dynamik von Dirac, Zeitschrift f\"{u}r Physik, Vol.53, No.3-4, 1929, pp157-165.

\bibitem[51]{51}
W. Greiner, {\it Relativistic Quantum Mechanics}, Springer-Verlag, 1990, pp261-267.

\bibitem[52]{52}
W. Greiner, B. M\"{u}ller and J. Rafelski, {\it Quantum Electrodynamics of Strong Fields}, Springer-
Verlag, 1985.


\end{thebibliography}
\end{document}